\documentclass[fleqn,usenatbib]{mnras}

\usepackage{newtxtext,newtxmath}

\usepackage[T1]{fontenc}

\DeclareRobustCommand{\VAN}[3]{#2}
\let\VANthebibliography\thebibliography
\def\thebibliography{\DeclareRobustCommand{\VAN}[3]{##3}\VANthebibliography}

\usepackage{ulem}

\usepackage{graphicx}	
\usepackage{amsmath}	

\usepackage{longtable}

\title[TangoSIDM and the Tully-Fisher relation]{TangoSIDM Project: Is the Stellar Mass Tully-Fisher relation consistent with SIDM?}
\author[C. A. Correa et al.]{Camila A. Correa$^{1}$\thanks{E-mail: camila.correa@cea.fr},
Matthieu Schaller$^{2,3}$, 
Joop Schaye$^3$, 
Sylvia Ploeckinger$^{2,4}$,
\newauthor
Josh Borrow$^{5}$, 
and Yannick Bahé$^{6,3,7}$
\\
$^1$ Universit\'e Paris-Saclay, Universit\'e Paris Cit\'e, CEA, CNRS, AIM, 91191, Gif-sur-Yvette, France \\
$^2$ Lorentz Institute for Theoretical Physics, Leiden University, PO Box 9506, NL-2300 RA Leiden, The Netherlands \\
$^3$ Leiden Observatory, Leiden University, PO Box 9513, NL-2300 RA Leiden, The Netherlands \\
$^4$ Department of Astropysics, University of Vienna, T\"urkenschanzstrasse 17, 1180 Vienna, Austria \\
$^5$ Department of Physics and Astronomy, University of Pennsylvania, 209 South 33rd Street, Philadelphia, PA, USA 19104 \\
$^6$ Institute for Computational Cosmology, Department of Physics, Durham University, South Road, Durham DH1 3LE, UK \\
$^7$ Laboratoire d’astrophysique, \'Ecole Polytechnique F\'ed\'erale de Lausanne (EPFL), 1290 Sauverny, Switzerland
}

\date{Accepted XXX. Received YYY; in original form ZZZ}

\pubyear{2024}

\begin{document}
\label{firstpage}
\pagerange{\pageref{firstpage}--\pageref{lastpage}}
\maketitle

\begin{abstract}
Self-interacting dark matter (SIDM) has the potential to significantly influence galaxy formation in comparison to the cold, collisionless dark matter paradigm (CDM), resulting in observable effects. This study aims to elucidate this influence and to demonstrate that the stellar mass Tully-Fisher relation imposes robust constraints on the parameter space of velocity-dependent SIDM models. We present a new set of cosmological hydrodynamical simulations that include the SIDM scheme from the TangoSIDM project and the SWIFT-EAGLE galaxy formation model. Two cosmological simulations suites were generated: one (Reference model) which yields good agreement with the observed $z=0$ galaxy stellar mass function, galaxy mass-size relation, and stellar-to-halo mass relation; and another (WeakStellarFB model) in which the stellar feedback is less efficient, particularly for Milky Way-like systems. Both galaxy formation models were simulated under four dark matter cosmologies: CDM, SIDM with two different velocity-dependent cross sections, and SIDM with a constant cross section. While SIDM does not modify global galaxy properties such as stellar masses and star formation rates, it does make the galaxies more extended. In Milky Way-like galaxies, where baryons dominate the central gravitational potential, SIDM thermalises, causing dark matter to accumulate in the central regions. This accumulation results in density profiles that are steeper than those produced in CDM from adiabatic contraction. The enhanced dark matter density in the central regions of galaxies causes a deviation in the slope of the Tully-Fisher relation, which significantly diverges from the observational data. In contrast, the Tully-Fisher relation derived from CDM models aligns well with observations. 
\end{abstract}



\section{Introduction}

The self-interacting dark matter paradigm (SIDM) postulates that dark matter particles engage in gravitational interactions with ordinary particles while exhibiting non-gravitational interactions among themselves. Arising as a natural prediction of dark sector models beyond the Standard Model (e.g. \citealt{Spergel00,Tulin18}), SIDM is expected to manifest detectable astrophysical signatures (e.g. \citealt{Adhikari22}). Moreover, it offers a potential explanation for the most challenging discrepancy between $\Lambda$ cold dark matter ($\Lambda$CDM) numerical simulations and observations: the diverse distribution of dark matter within dwarf galaxies (see e.g. \citealt{Oman15,Santos20,Hayashi21,Sales22,Borukhovetskaya22}).

Within the SIDM framework, interactions among dark matter particles dynamically alter the internal structure of dark matter halos. This modification involves the transfer of heat from the outer parts to the inner halo, resulting in an increase in the velocity dispersion, and a reduction of dark matter densities in the central regions (e.g. \citealt{Dave01,Colin02,Vogelsberger12,Rocha13,Dooley16,Vogelsberger16}). The crucial parameter governing the rate of dark matter particle interactions is the cross section per unit mass, denoted as $\sigma/m_{\chi}$ (e.g. \citealt{Robertson17,Kahlhoefer19,Kummer19,Vogelsberger19,Banerjee20,Shen21}). Measurements derived from the shape and collision of nearby galaxy clusters constrain this parameter to be ${<}1~\rm{cm}^2\rm{g}^{-1}$ (e.g. \citealt{Randall08,Dawson13,Massey15,Harvey15,Wittman18,Harvey19,Sagunski21,Andrade22}). 

While various studies have explored the impact of SIDM under a small and constant cross section, prevailing particle physics models advocate for a velocity-dependent framework, where $\sigma/m_{\chi}$ allows dark matter to behave as a collisional fluid on small scales while remaining essentially collisionless over large scales (e.g. \citealt{Pospelov08, ArkaniHamed09,Buckley10,Feng10,Boddy14,Tulin18}). Under this velocity-dependent scheme, $\sigma/m_{\chi}$ can be ${<}1~\rm{cm}^2\rm{g}^{-1}$ for high dark matter velocities at large scales, aligning with the constrains of cluster-size haloes, and exceed ${>}100~\rm{cm}^2\rm{g}^{-1}$ for low dark matter velocities in order to explain the diverse dark matter distribution within dwarf galaxies (e.g. \citealt{Correa21,Gilman21,Correa22,Yang23,Silverman23,Nadler23,Shah23,Gilman23}). Although SIDM has been robustly constrained on galaxy cluster scales, uncertainties persist in the lower-mass galaxy regime due to the difficulty in isolating the impact of baryonic physics from dark matter interactions.

Recent studies exploring the co-evolution of baryons and SIDM in isolated systems indicate that non-bursty stellar feedback may not significantly alter SIDM density profiles in dwarf galaxies (e.g. \citealt{Vogelsberger14, Robles17, Sameie21}). Conversely, hydrodynamical simulations incorporating SIDM and a bursty stellar feedback model reveal distinctions in velocity dispersion profiles between SIDM and CDM haloes (\citealt{Burger22}), suggesting the need for more detailed investigations into the interplay between SIDM and various feedback models. In more massive systems, the intricate interplay between SIDM and baryons is even more challenging. Studies that modelled the evolution of Milky Way-like systems and galaxy clusters (e.g. \citealt{Robertson19,Despali19,Sameie21,Rose22}) found that baryon contraction results in the formation of denser and cuspier central density profiles under SIDM compared to CDM. Analytical studies focusing on the gravitational contribution of a baryonic disc and bulge reached similar conclusions (\citealt{Robles19, Silverman23, Jiang23}). However, uncertainties persist regarding how the increased cuspiness of SIDM haloes depends on the specific SIDM model parameters or the strength of galaxy feedback models.

This paper seeks to address this knowledge gap by introducing a new set of cosmological hydrodynamical simulations. These simulations integrate the SIDM model derived from the TangoSIDM project, with the baryonic physics from the SWIFT-EAGLE galaxy formation model. The goals of the TangoSIDM project are to derive robust constraints on the dark matter cross section from observations of dwarf and Milky Way-type galaxies. In this study, we take a pivotal first step by demonstrating how the stellar mass Tully-Fisher relation, a well-established galaxy scaling relation, can be leveraged to derive robust constraints on the parameter space of velocity-dependent SIDM models. The structure of this paper is organized as follows. Section \ref{Model_section} describes the SIDM and baryonic subgrid models employed in our simulations. In Section \ref{Gal_prop_section}, we show how SIDM influences key galaxy properties, including stellar masses, sizes, and star formation rates. Section \ref{Density_sec} compares the dark matter density profiles of haloes between CDM and various SIDM models. Section \ref{TFRelation_sec} undertakes an in-depth analysis of the stellar mass Tully-Fisher relation, and demonstrates it rules out the velocity-dependent SIDM models studied in this work. Section~\ref{Discussion_sec} discusses the SIDM parameter space, and Section \ref{Conclusion_sec} summarizes the paper's findings.

\section{Simulation setup}\label{Model_section}

TangoSIDM\footnote{www.tangosidm.com} is a simulation project dedicated to modelling cosmological simulations that capture the intricacies of structure formation within a $\Lambda$SIDM universe. This work introduces the first realization of hydrodynamical cosmological volumes, each spanning 25 Mpc on a side, as integral compontents of the TangoSIDM project. To produce these simulations, the SWIFT\footnote{www.swiftsim.com} code (\citealt{Schaller23}) was employed. SWIFT includes advanced hydrodynamics and gravity schemes. The gravity solver employs the Fast Multiple Method (\citealt{Greengard87}) with an adaptive opening angle, while for hydrodynamics the SPHENIX SPH scheme (\citealt{Borrow22}), specifically designed for galaxy formation sub-grid models, was utilized.

The simulations follow the evolution of 376$^{3}$ dark matter particles and 376$^{3}$ gas particles to redshift $z=0$. The softening is set to 2.66 comoving kpc at early times, but is frozen a physical value of 700 pc at $z=2.8$. The dark matter particle mass is $9.70\times 10^{6}~\rm{M}_{\odot}$ and the gas initial particle mass is $1.81\times 10^{6}~\rm{M}_{\odot}$. The starting redshift of the simulations is $z=127$. The initial conditions were calculated using second-order Lagrangian perturbation theory with the method of \citet{Jenkins10,Jenkins13}. The adopted cosmological parameters are $\Omega_{\rm{m}}=0.307$, $\Omega_{\Lambda}=0.693$, $h=0.6777$, $\sigma_{8}=0.8288$ and $n_{\rm{s}}=0.9611$ (\citealt{Planck13}).

\subsection{TangoSIDM model}

\begin{figure} 
	\includegraphics[angle=0,width=0.45\textwidth]{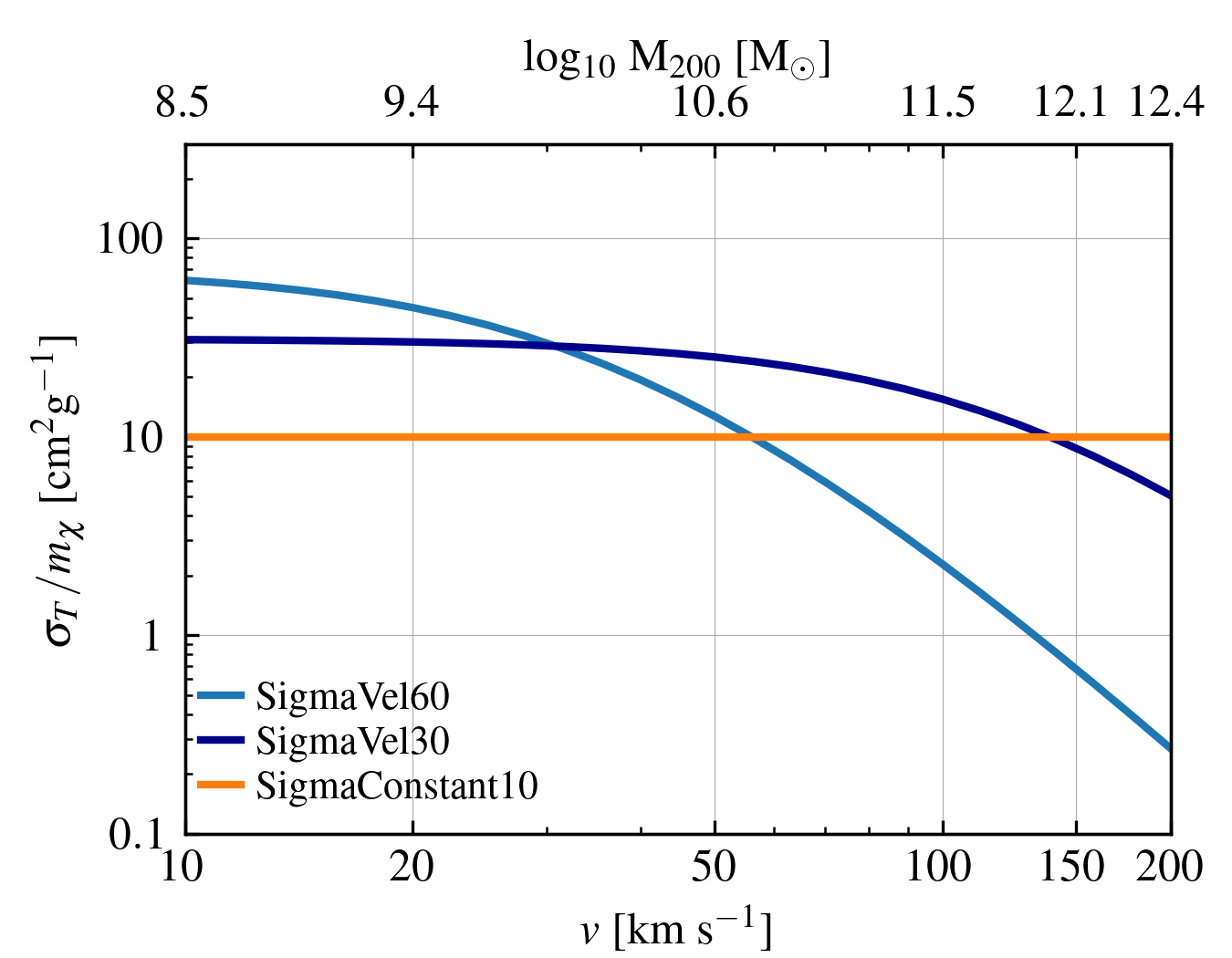}
	\caption{Momentum transfer cross section as a function of the relative scattering velocity among dark matter particles for the SIDM models featured in this work (Table~\ref{Table_sidm_models}). The figure shows two velocity-dependent models, namely SigmaVel60 (light blue line) and SigmaVel30 (dark blue line), alongside SigmaConstant10 (orange line), which uses a constant cross section, $\sigma_{T}/m_{\chi}=10$ cm$^{2}$g$^{-1}$. The top x-axis indicates the typical halo mass that hosts orbits of the velocities indicated on the bottom x-axis.}
	\label{sidm_models}
\end{figure}

The TangoSIDM project, encompassing its models and SIDM implementation, was presented in \citet{Correa22}. In this section we briefly summarise the key elements of the SIDM model, with further details available in the aforementioned reference.

Four dark matter models were generated for this study: the cold collisionless dark matter model (hereafter CDM); a SIDM model with a constant scattering cross section of 10 cm$^{2}\rm{g}^{-1}$ (hereafter SigmaConstant10); and two SIDM models featuring velocity-dependent cross sections (see Fig.~\ref{sidm_models}). Although the SigmaConstant10 model has been ruled out by observations of galaxy clusters (e.g.~\citealt{Harvey15,Harvey19}), it serves as a control model for comparative analysis. Among the velocity-dependent models, one has a cross section that is below 1 cm$^{2}\rm{g}^{-1}$ at high velocities ($v{>}150$ km s$^{-1}$) and increases with decreasing velocity, reaching 60 cm$^{2}\rm{g}^{-1}$ at 10 km s$^{-1}$ (hereafter SigmaVel60 model). The other velocity-dependent model has a cross section smaller than 8 cm$^{2}\rm{g}^{-1}$ at velocities surpassing $200$ km s$^{-1}$ (dropping below 1 cm$^{2}\rm{g}^{-1}$ at ${\approx}1000$ km s$^{-1}$) and increases with decreasing velocity, reaching 30 cm$^{2}\rm{g}^{-1}$ at 10 km s$^{-1}$ (hereafter SigmaVel30 model).

The SigmaVel60 and SigmaVel30 models represent two extreme scenarios for the rate of dark matter interactions in Milky Way-mass systems. Despite both models adhering to the SIDM constraints derived from cluster-size haloes, there are important differences. In SigmaVel60, interactions reach 1-2 cm$^{2}\rm{g}^{-1}$ around 100 km s$^{-1}$, therefore this model produces a low rate of interactions in the center of Milky Way-like haloes. In contrast, SigmaVel30 exhibits a cross section of 10-20 cm$^{2}\rm{g}^{-1}$ at 100 km s$^{-1}$, imposing a stronger rate of interaction.

The velocity-dependent cross sections are modelled under the assumption that dark matter particle interactions are mediated by a Yukawa potential dependent on three parameters: the dark matter mass $m_{\chi}$; the mediator mass $m_{\phi}$; and the coupling strength $\alpha_{\chi}$. While there is no analytical form for the differential scattering cross-section due to a Yukawa potential, the Born-approximation (\citealt{Ibe10})$-$applicable when treating the scattering potential as a small perturbation$-$yields the differential cross-section of the dark matter-dark matter interactions

\begin{equation}\label{sigmam}
\frac{{\rm{d}}\sigma}{{\rm{d}}\Omega}=\frac{\alpha_{\chi}^{2}}{m_{\chi}^{2}(m_{\phi}^{2}/m_{\chi}^{2}+v^{2}\sin^{2}(\theta/2))^{2}}.
\end{equation}

While in the model with a constant cross section the dark matter scattering is isotropic, in the velocity-dependent cross section models the scattering is anisotropic. For anisotropic scattering the momentum transfer cross section, defined as

\begin{equation}\label{sigmat}
\sigma_{T}/m_{\chi}=2\int (1-|\cos\theta|)\frac{{\rm{d}}\sigma}{{\rm{d}}\Omega}{\rm{d}}\Omega,
\end{equation}

\noindent is useful to consider, because it is weighted by the scattering angle and therefore it does not overestimate the scattering with $\theta>\pi/2$ (\citealt{Kahlhoefer15}). Table~\ref{Table_sidm_models} shows the SIDM model parameters adopted in this work and Fig.~\ref{sidm_models} shows the momentum transfer cross sections. The figure shows the velocity-dependent models (light blue and dark blue lines) and the constant cross section model (orange line). While the bottom x-axis shows the relative velocity between dark matter particles, the top x-axis indicates the typical halo mass that hosts circular orbits of such velocities. 

\begin{table}
\begin{center}
\caption{SIDM models analysed in this work. Form left to right: Model name, SIDM parameters for each model (dark matter mass, $m_{\chi}$, mediator mass, $m_{\phi}$, and coupling strength, $\alpha$) and type of dark matter interaction.}
\begin{tabular}{lcccl}
\hline
\multicolumn{1}{c}{} & \multicolumn{3}{c}{\uline{SIDM parameters}} & \multicolumn{1}{c}{\uline{DM interaction}} \\
Model & $m_{\chi}$ & $m_{\phi}$ & $\alpha$ &\\
Name & [GeV] & [MeV] &  & \\
\hline
CDM & - & - & - & No interaction\\
SigmaConstant10 & - & - & - & Isotropic\\
SigmaVel30 & 2.227 & 0.778 & $4.317\times 10^{-5}$ & Anisotropic\\
SigmaVel60 & 3.855 & 0.356 & $1.027\times 10^{-5}$ & Anisotropic\\
\hline
\end{tabular}
\end{center}
\label{Table_sidm_models}
\end{table}

\subsection{SWIFT-EAGLE model}

The SWIFT-EAGLE model, an open-source galaxy formation model implemented in SWIFT, is derived from the original EAGLE model (\citealt{Schaye15,Crain15}). While it has common modules to those of EAGLE, SWIFT-EAGLE includes new developments and improvements. A detailed model description can be found in \citet{Bahe22} and \citet{Borrow22b}. Below we provide a summary.

SWIFT-EAGLE incorporates the element-by-element sub-grid radiative gas cooling and photoheating prescription from \citet{Ploeckinger20}, which accounts for the inter-stellar radiation field and self-shielding of dense gas, as well as the UV/X-ray background from galaxies and quasars according to \citet{FaucherGiguere20}. Star formation is implemented stochastically, following the \citet{Schaye08} pressure law, as in the original EAGLE model. A polytropic equation of state, $P\propto \rho^{4/3}$, sets a minimum limit on the gas pressure. The star formation rate per unit mass is calculated from the gas pressure, employing an analytical formula designed to reproduce the observed Kennicutt–Schmidt law (\citealt{Kennicutt98}) in disc galaxies. A gas particle is star-forming if its subgrid temperature $T<1000$ K, or if its density (expressed in units of hydrogen particles per cubic cm, $n_{\rm{H}}$) is $n_{\rm{H}}>10~\rm{cm}^{-3}$ and temperature $T<10^{4.5}$ K.

The stellar initial mass function assumes the form of \citet{Chabrier03} within the range 0.1-100 M$_{\odot}$, with each particle representing a simple age stellar population. Stellar feedback is implemented stochastically, following the prescription of \citet{DallaVecchia12}, where stars with masses between 8 $\rm{M}_{\odot}$ and 100 $\rm{M}_{\odot}$ explode as core-collapse supernovae. The resulting energy is transferred as heat to the surrounding gas, following \citet{Chaikin22}.

The energy injected into the gas corresponds to $10^{51}$ erg per supernova times a dimensionless coupling efficiency factor, $f_{\rm{E}}$, that follows the same scaling function as in EAGLE,

\begin{equation}\label{feedback}
f_{\rm{E}}=f_{\rm{E,max}}-\frac{f_{\rm{E,max}}-f_{\rm{E,min}}}{1+\rm{exp}\left(\frac{-\log_{10}Z/Z_{0}}{\sigma_{Z}}\right) \rm{exp}\left(\frac{\log_{10}n_{H}/n_{H,0}}{\sigma_{n}}\right) }.
\end{equation}

\noindent As can be seen, $f_{\rm{E}}$ depends on a number of free parameters: $f_{\rm{E,min}}$ and $f_{\rm{E,max}}$, which set the minimal and maximal feedback energies, $n_{\rm{H},0}$ and $Z_{0}$ defined as the density and metallicity pivot point around which the feedback energy fraction plane rotates, and $\sigma_{Z}$ and $\sigma_{n}$, the width of the feedback energy fraction sigmoids in the metallicity and density dimensions.

In addition to the energy released through star formation, star particles also release metals into the inter-stellar medium (ISM) through four evolutionary channels: AGB stars, winds from massive stars, core-collapse supernvae and Type Ia supernovae. This process follows the methodology discussed in \citet{Wiersma09} and \citet{Schaye15}. The abundances of 9 elements (H, He, C, N, O, Ne, Mg, Si, Fe) are tracked.

The formation and growth of supermassive black holes are modelled following \citet{Bahe22}. Initially seeded within friends-of-friends dark matter groups of mass $10^{10}$ M$_{\odot}$, black holes accretion rates follow the Eddington-limited Bondi accretion rate. The feedback mechanism from active galactic nucleus (AGN) activity is implemented following \citet{Booth09}. The energy depends on the accreted mass, $\Delta m$, onto the black hole as, $\Delta E=\epsilon_{\rm{r}}\epsilon_{\rm{f}}\Delta m c^{2}$, where $\epsilon_{\rm{r}}=0.1$ is the default value. This energy is stored in a reservoir carried by each black hole particle until it can be utilized to heat the nearest gas particle, inducing a temperature increase of $\Delta T_{\rm{AGN}}$. The coupling efficiency, $\epsilon_{\rm{f}}$, and the heating temperature of AGN feedback are free parameters.

\subsection{Reference \& WeakStellarFB SWIFT-EAGLE models}

This work investigates the evolution of galaxies for two distinct SWIFT-EAGLE models. In the first, referred to as the Reference model, the free parameters described in the previous subsection were calibrated in a (25 Mpc)$^3$ volume to reproduce the galaxy stellar mass function and galaxy mass-size relation. The second, named the WeakStellarFB model, adopts parameters that produce Milky Way-mass galaxies with very weak stellar feedback. Table~\ref{Table_eagle_models} provides a comprehensive listing of the subgrid parameter values for both models. 

The parameters for the Reference model were derived within the CDM framework using emulators that employed the Gaussian Process Regression-based python module SWIFTEmulator (\citealt{Kugel22}). Further details on the calibration and emulation technique can be found in \citet{Borrow22b}. Note that the SIDM simulations with the SWIFT-EAGLE Reference model adopt the parameters listed in Table~\ref{Table_eagle_models}, no re-calibration was performed to account for the SIDM effects.

The original parameters from the EAGLE simulations were calibrated to reproduce the $z=0.1$ galaxy stellar mass function, the relation between galaxies stellar mass and galaxies' central black hole masses, as well as disc galaxy sizes (\citealt{Crain15}). While the SWIFT-EAGLE model was inspired by EAGLE, significant differences exist, such as the gravity and hydrodynamics solver, cooling rates, supernovae and AGN feedback energy deposition into the ISM. Because of these differences, applying the original EAGLE parameter values in the SWIFT-EAGLE model yields different results. Relative to the Reference model, the WeakStellarFB model exhibits a weaker stellar feedback at the specific mass scale of $10^{12}~\rm{M}_{\odot}$ haloes, attributed to the lower value of $f_{\rm{E,max}}$ and higher $n_{\rm{H,0}}$. This combination results in a lower coupling efficiency factor $f_{\rm{E}}$ at fixed hydrogen number density, justifying its nomenclature ``WeakStellarFB''. 

In Section~\ref{Gal_prop_section} and Appendix~\ref{AppendixA}, we show that both the Reference and WeakStellarFB models yield stellar mass functions, specific star formation rates, and stellar-to-halo mass relations that closely align with observational data. However, the stellar feedback in the WeakStellarFB model is less efficient in Milky Way-mass systems, making them more compact by redshift zero. The primary objective of exploring SIDM under these two galaxy models is to understand the impact of dark matter collisions in the central regions of galaxies. We aim to discern how SIDM coevolves with the dynamical heating from supernova explosions and evaluate whether our conclusions regarding the impact of SIDM on galaxies remain robust in the face of variations in feedback models.

\begin{table}
\begin{center}
\caption{Subgrid parameter values of the SWIFT-EAGLE galaxy formation model that regulate stellar and AGN feedback. The left column identifies each parameter, with detailed descriptions provided in the text. The middle and right columns list the parameter values adopted in the Reference and WeakStellarFB models, respectively.}
\begin{tabular}{lll}
\hline
Parameters & Reference & WeakStellarFB \\
\hline
$f_{\rm{E,min}}$ & 0.388 & 0.5\\
$f_{\rm{E,max}}$ & 7.37 & 5.0\\
$n_{\rm{H,0}}$ [cm$^{-3}$] & 0.412 & 1.46\\
$\sigma_{\rm{Z}}$ & 0.311 & 0.275\\
$Z_{0}$ & 0.00134 & 0.00134 \\
$\sigma_{\rm{n}}$ & 0.428 & 1.77\\
$\epsilon_{\rm{f}}$ & 0.035 & 0.1\\
$\Delta T_{\rm{AGN}}$ [K] & 10$^{8.62}$ &10$^{8.5}$ \\
\hline
\end{tabular}
\end{center}
\label{Table_eagle_models}
\end{table}

\begin{figure*} 
	\includegraphics[angle=0,width=\textwidth]{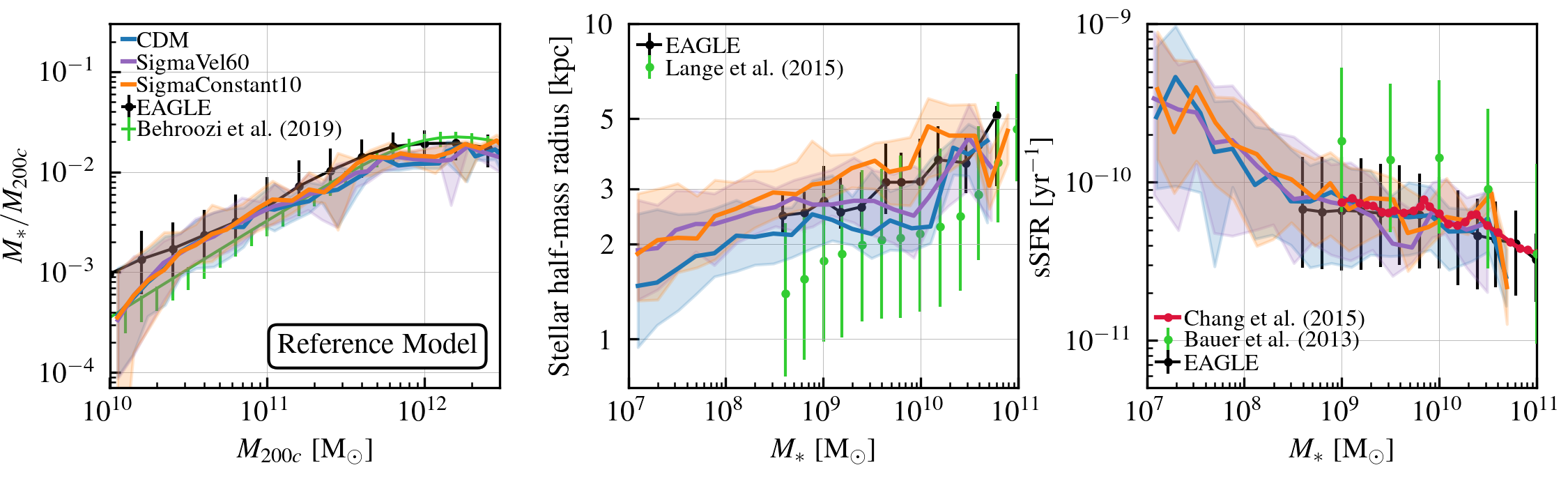}
	\includegraphics[angle=0,width=\textwidth]{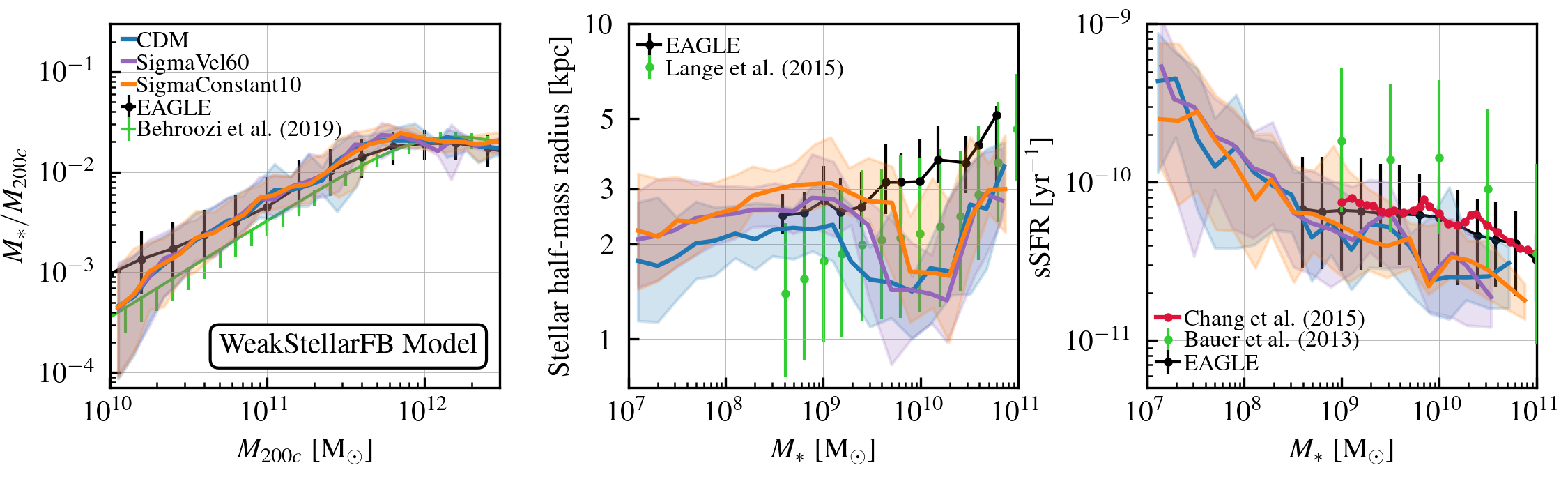}
	\caption{Galaxy scaling relations at redshift $z=0$ for the Reference (top panels) and WeakStellarFB model (bottom panels). The columns show the stellar-to-halo mass ratio ($M_{*}/M_{200\rm{c}}$) as a function of halo mass (left), the projected stellar half-mass radius as a function of stellar mass (middle) for all galaxies, and the specific star formation rate (sSFR, SFR/$M_{*}$) as a function of stellar mass for actively star forming-galaxies (right). In all panels the curves correspond to the median for the Reference and WeakStellarFB models produced in CDM (blue lines), SigmaVel60 (purple lines) and SigmaConstant10 (orange lines) frameworks. SigmaVel30, though not shown, follows a similar trend as SigmaVel60. The shaded regions mark the 16th-84th percentiles of the relations. These models are contrasted with various observational datasets and the EAGLE simulations (black lines). The left panels show the stellar-to-halo mass relation from \citet{Behroozi19}. In the middle panels, galaxy sizes are compared with citet{Lange15} (green circles) dataset. The right panels compare the sSFR with those reported by \citet{Bauer13} and \citet{Chang15}. SIDM appears to have minimal impact on galaxy masses and star formation rates. However, it significantly alters galaxy sizes, leading to increases by up to a factor of 2 for SigmaConstant10.}
	\label{GalProp_fig}
\end{figure*}

\subsection{Halo catalogue and definitions}

Halo catalogues were generated using the VELOCIraptor halo finder (\citealt{Elahi11,Elahi19,Canas19}). VELOCIraptor uses a 3D-friends of friends (FOF) algorithm to identify field haloes, and subsequently applies a 6D-FOF algorithm to separate virialised structures and identify sub-haloes of the parent haloes (\citealt{Elahi19}). Throughout this work, virial halo masses ($M_{200\rm{c}}$) are defined as all matter within the virial radius $R_{200\rm{c}}$, for which the mean internal density is 200 times the critical density, $\rho_{\rm{crit}}$, which is $127.5\rm{M}_{\odot}\rm{kpc}^{-3}$ at $z=0$. In each FOF halo, the `central' subhalo is the one that is most likely the core in phase space, which is nearly always the most massive. The remaining subhaloes within the FOF halo are its satellites. The resolution of the simulations is sufficient to resolve (sub-)haloes down to ${\sim}10^{10}~\rm{M}_{\odot}$ with $10^{3}$ particles within $R_{200}$. Galaxy stellar masses, sizes and star formation rates are always defined within an aperture of 50 kpc.

\section{Galaxy properties}\label{Gal_prop_section}

In this section we analyse key galaxy properties from the Reference and WeakStellarFB models: the $z=0$ stellar-to-halo mass relation, projected galaxy sizes, and star formation rates, and we compare them against observational data. It is important to point out that during the calibration of the subgrid parameters for feedback under CDM, the $z=0$ galaxy stellar mass function and the stellar mass-size relation were considered, and as a result, the simulations do not provide predictions for these. We remind the reader that the subgrid parameters from the Reference model were only calibrated under the CDM framework and not under SIDM. The SIDM simulations use the same subgrid parameter values as CDM for both the Reference and WeakStellarFB models. The $z{=}0$ galaxy stellar mass function is presented in Appendix~\ref{AppendixA}.

Fig.~\ref{GalProp_fig} illustrates three galaxy scaling relations from the Reference (the top panels) and WeakStellarFB models (bottom panels). In the left panels, the ratio between the galaxy stellar mass and halo mass ($M_{*}/M_{200\rm{c}}$) is plotted as a function of the host halo mass. Coloured curves represent the median relations for central galaxies, with shaded regions indicating the 16-84th percentiles. A comparison is made with the stellar-to-halo mass relation from the EAGLE simulation and from UNIVERSEMACHINE (\citealt{Behroozi19}). Notably, the WeakStellarFB model aligns best with the original EAGLE data (\citealt{McAlpine16}). At fixed halo mass, galaxies from the WeakStellarFB model are more massive than those from the Reference model (consistent with a comparison of the stellar mass functions). The dark matter framework does not significantly alter the stellar-to-halo mass relation. For clarity, the SigmaVel30 model is not shown, as it follows a trend similar to SigmaVel60.

Moving to the middle panels of Fig.~\ref{GalProp_fig}, the stellar half-mass radius is shown as a function of stellar mass. The half-mass radius is defined as the radius that encloses 50 per cent of the stellar mass, and is computed from all bound star particles within a projected 2D circular aperture of 50 kpc radius. The simulations are compared against the GAMA survey (\citealt{Lange15}), and the EAGLE simulation (\citealt{McAlpine16}). An interesting feature emerges in the bottom middle panel, revealing a U-shape trend in the galaxy size-mass relation. Galaxies within the mass range of $10^{9}$ to $10^{11}~\rm{M}_{\odot}$ become too compact due to excessive radiative losses at high gas densities. To counteract this issue, the EAGLE model introduced a dependence of the stellar feedback energy on the gas density (eq.~\ref{feedback}), so that higher density gas receives a larger amount of energy from stellar explosions (\citealt{Crain15}). The WeakStellarFB model incorporates the density-dependent stellar feedback energy, but its parameter values are such that the feedback strength remains inadequate. The coupling efficiency factor applied to the supernova energy that is injected into that gas is smaller than in the Reference model. Therefore, while stellar and AGN feedback in the WeakStellarFB model can prevent the formation of excessively massive galaxies, it does not guarantee the formation of extended galaxies with realistic sizes. A more careful approach, or tuning of the energy parameters, is required for feedback to effectively eject low-angular momentum gas, increase the median angular momentum of the ISM gas that remains to form stars, and form more extended galaxies (e.g. \citealt{Brook12}). 

For stellar masses ${\approx}10^{9}~\rm{M}_{\odot}$ the WeakStellarFB model predicts galaxies with sizes that agree with EAGLE, and do not seem to suffer from overcooling and compactness. However, these sizes appear large when compared to the dataset of \citet{Lange15}. The Reference model, calibrated to match the size-mass relation from \citet{Lange15}, yields galaxies that are still overly extended, partly due to the sampling noise in gravitational interactions between stars and dark matter, that leads to spurious size growth (\citealt{Ludlow19,Ludlow23}).
 
The middle panels also show the evident impact of SIDM on galaxy sizes. Dark matter particle interactions heat the inner halo, leading to core formation in the central regions and dynamically heating the surrounding gas and stars, promoting the formation of more extended galaxies. However, this is insufficient to counteract the overcooling and compactness observed in the WeakStellarFB model for galaxies more massive than $10^{10}~\rm{M}_{\odot}$. The top middle panel shows that the SigmaVel60 model, characterized by a large cross section for galaxies less massive than $10^9M_{\odot}$, produces sizes that are close to those for the SigmaConstant10 model. For these masses, as the cross section decreases, the galaxy sizes from SigmaVel60 decrease relative to those for SigmaConstant10, and become similar to those of CDM.

\citet{Ludlow23} found than in CDM hydrodynamical simulations like EAGLE, which share the same numerical resolution as the TangoSIDM simulations, the galaxies' half-mass radius remains robust against spurious collisional heating only for halo masses $M_{200\rm{c}}{\gtrsim}10^{11.7}~\rm{M}_{\odot}$. This suggests that our galaxies' sizes are free from spurious heating if the galaxies are more massive than $M_{*}{\gtrsim}10^{10}~\rm{M}_{\odot}$. We note, however, that resolution effects may have a stronger impact on CDM simulations than on SIDM simulations, in which case the effect of SIDM on sizes relative to CDM may be underestimated.

The right panels of Fig.~\ref{GalProp_fig} display the median specific star formation rates (sSFR) for actively star forming-galaxies, with galaxies classified as star-forming if their sSFR ${>}10^{-11}$ yr$^{-1}$. The panels reveal that the $z=0$ sSFR from the Reference model are in agreement with the sSFR from the EAGLE simulations and the dataset from \citet{Chang15}, and are within a factor of 5 from the \citet{Bauer13} data. The WeakStellarFB model has sSFR lower than Reference. Interestingly, there are no differences in the median sSFR trends between simulations with CDM vs. SIDM.

\begin{figure*} 
\includegraphics[angle=0,width=0.9\textwidth]{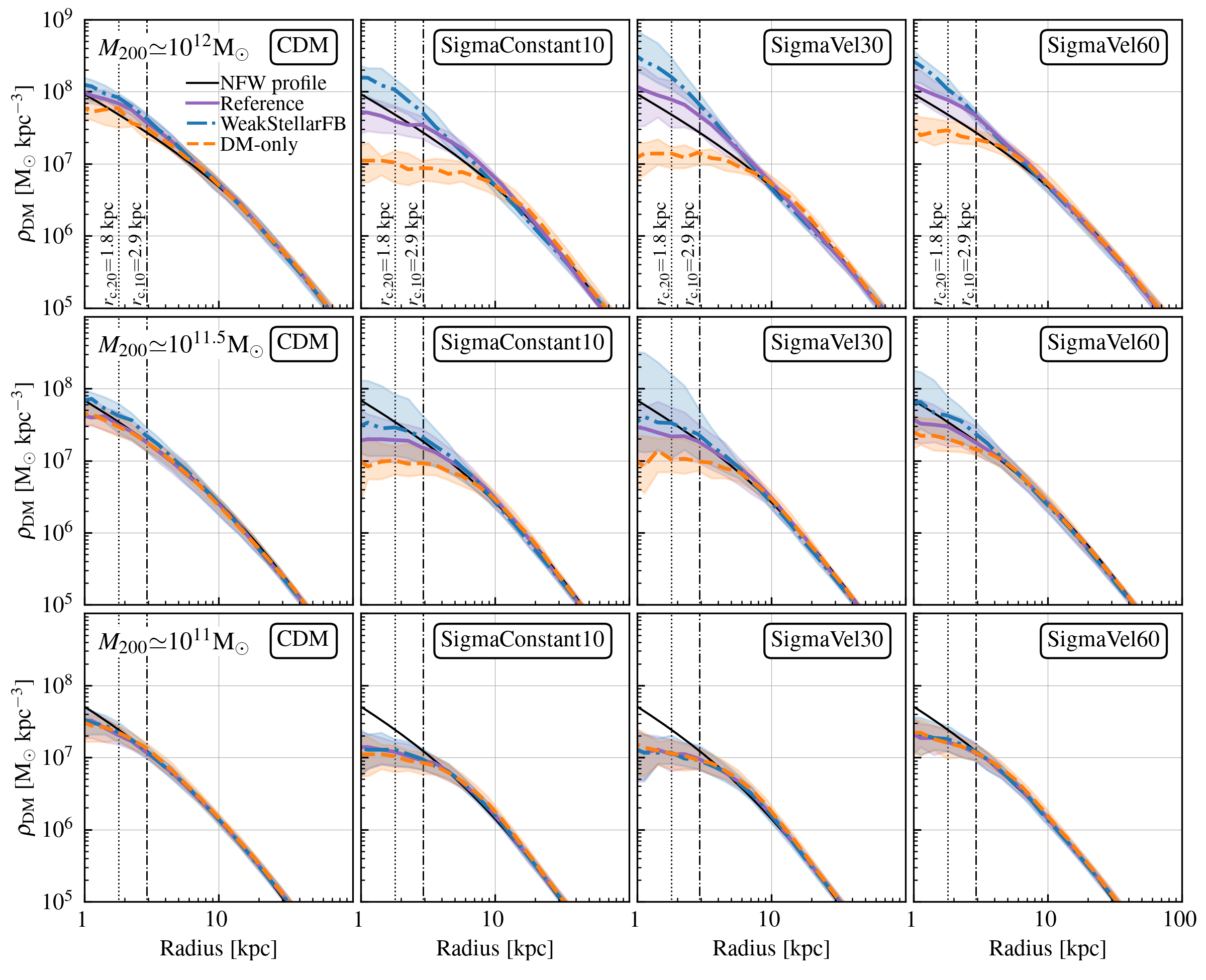}
\caption{Dark matter density profiles, $\rho_{\rm{DM}}$, of $10^{11}$ M$_{\odot}$, $10^{11.5}$ M$_{\odot}$ and $10^{12}$ M$_{\odot}$ haloes from the Reference (purple solid lines) and WeakStellarFB (blue dot-dashed lines) models under CDM (left panels), SigmaConstant10 (second panels from the left), SigmaVel30 (third panels from the left) and SigmaVel60 (right panels). The panels compare $\rho_{\rm{DM}}$ between hydrodynamical (CDM and SIDM) simulations and dark matter-only simulations (orange dashed lines). The coloured lines highlight the median values and the shaded regions the 16-84th percentiles. Additionally, the black solid line corresponds to the NFW profile (estimated using the concentration-mass relation from \citealt{Correa15c}), and the black dotted and dashed-dotted lines indicate the convergence radii (see text for definition). The differences in the profiles between haloes of the same mass highlights the impact of baryonic effects and dark matter particle interactions on the central haloes densities.}
\label{Density_profiles}
\end{figure*}

\begin{figure} 
\includegraphics[angle=0,width=0.49\textwidth]{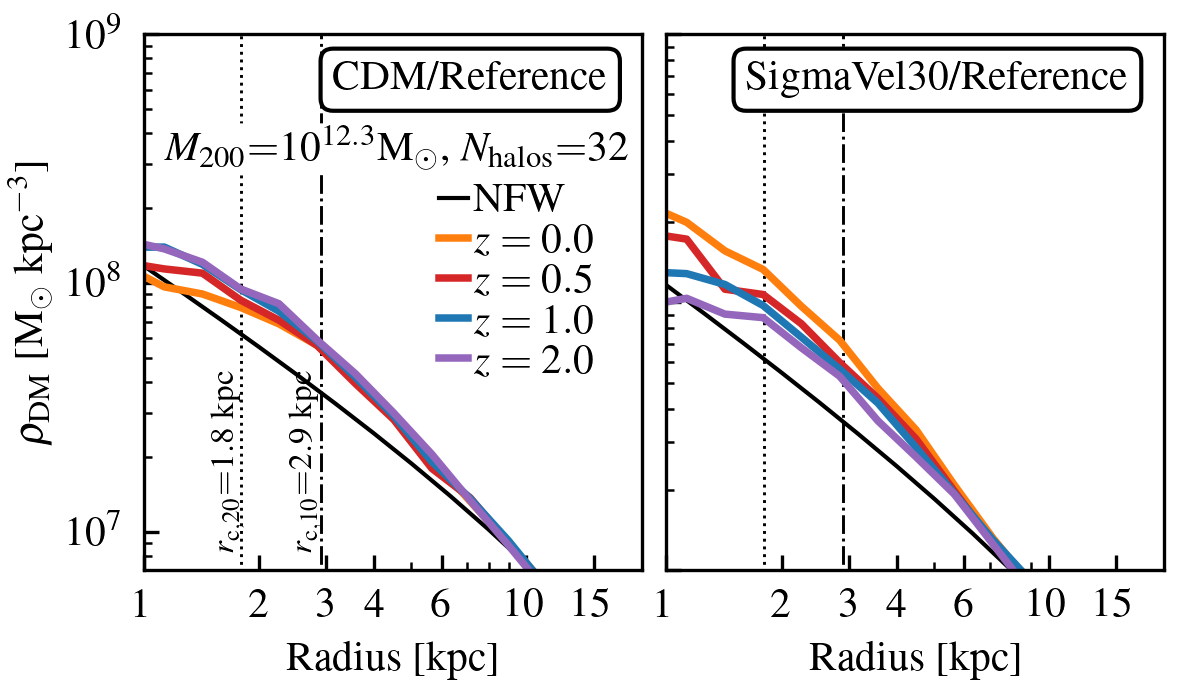}
\caption{Stacked dark matter density profiles, $\rho_{\rm{DM}}$, of the 32 most massive haloes in the box (with masses larger than $10^{12}$ M$_{\odot}$) at $z=0$ from the Reference model under CDM (left panel) and SigmaVel30 (right panel). The panels show the median density evolution between redshifts 0 and 2. The coloured lines highlight the median values and the black solid line shows the NFW profile of the haloes at redshift zero (estimated using the concentration-mass relation from \citealt{Correa15c}). The black dotted and dashed-dotted lines indicate the convergence radii (see text for definition). While there is no large difference in the median density profiles of haloes over time in the CDM, the SigmaVel30 model shows that the central density increases.}
\label{Density_evolution_1}
\end{figure}

\section{Dark Matter Density profile}\label{Density_sec}

In the following analysis, we compare our findings with prior studies on SIDM. Specifically, we examine the dark matter density profiles of central haloes with masses in the range of $10^{10.9}-10^{11.1}$ M$_{\odot}$, $10^{11.4}-10^{11.6}$ M$_{\odot}$ and $10^{11.9}-10^{12.1}$ M$_{\odot}$ from both the Reference and WeakStellarFB models under CDM, SigmaConstant10, SigmaVel30 and SigmaVel60.

The panels in Fig.~\ref{Density_profiles} compare the dark matter density profiles, $\rho_{\rm{DM}}$, between hydrodynamical simulations (CDM and SIDM) of the Reference model (purple solid lines) and the WeakStellarFB model (blue dot-dashed lines). Additionally, dark matter-only simulations are presented as orange dashed lines. To facilitate the comparison, the NFW density profile (black solid lines) is included, estimated using the concentration-mass relation from \citet{Correa15c}. We also include convergence radii defined as the minimum radius where the mean density converges at the 20 and 10 per cent level, $r_{\rm{c},20}$ (dash-dotted lines) and $r_{\rm{c},10}$ (dotted lines) respectively, relative to a simulation of higher resolution. The convergence criterion $r_{\rm{c},10}$, presented by \citet{Ludlow19b}, is defined as $r_{\rm{c},10}=0.055 l(z)$, where $l(z)$ is the (comoving) mean inter-particle separation. At $z=0$ this separation is $l=L_{\rm{b}}/N_{\rm{p}}^{1/3}=52.7$ kpc, given $L_{\rm{b}}=25$ cMpc and $N_{\rm{p}}=2\times 376^{3}$ particles. In addition to \citet{Ludlow19b} criterion, we include a relaxed convergence criterion given by $r_{\rm{c},20}=0.034 l(z)$. This is motivated by the findings of \citet{Schaller15}, who showed that the differences in the mean density profiles from the EAGLE hydrodynamical and DM-only simulations are significantly larger than 10\%. The value of 0.034 is obtained from eq. (18) of \citet{Ludlow19b} after decreasing $\kappa_{\rm{P}03}\equiv \frac{t_{\rm{relax}}}{t_{200}}$ by a factor of 2 (see \citealt{Power03}).

The bottom panels of Fig.~\ref{Density_profiles} show that, in $10^{11}$ M$_{\odot}$ haloes, baryons do not affect $\rho_{\rm{DM}}(r)$ beyond $r_{\rm{c},20}$. Under both CDM and SIDM, the hydrodynamical and DM-only simulations yield consistent $\rho_{\rm{DM}}$. In the CDM models, $\rho_{\rm{DM}}(r)$ agrees with the NFW prediction for $r>r_{\rm{c},20}$, while in SIDM models, dark matter particle interactions create the expected constant-density isothermal cores (see also e.g., \citealt{Colin02,Vogelsberger12,Peter13,Rocha13}, and \citealt{Correa22} for DM-only TangoSIDM density profiles). This cored $\rho_{\rm{DM}}$ corresponds to the median profile of the central $10^{11}$ M$_{\odot}$ halo population. However, note that since the velocity-dependent SIDM models under consideration exhibit large cross sections at the $10^{11}~\rm{M}_{\odot}$ mass-scale, some SIDM haloes may potentially undergo core-collapse and form a cuspy central density profile.

The bottom panels of Fig.~\ref{Density_profiles} reveal that galaxies with stellar masses as high as $10^{9}$ M$_{\odot}$ do not produce sufficiently strong feedback to affect the underlying dark matter distribution. In agreement with our results, \citet{Robles17} modelled dwarf galaxies within $10^{10}~\rm{M}_{\odot}$ haloes under both CDM and SIDM using the zoom-in FIRE cosmological model. They concluded that, for these low-mass systems, the final density profile of SIDM haloes was not strongly influenced by the stellar mass of the galaxy, exhibiting cored density profiles regardless of hosting galaxies with stellar masses ranging from $10^5$ to $10^7~\rm{M}_{\odot}$. Furthermore, \citet{Burger22} showed that both CDM and SIDM can yield haloes with cored density profiles. The difference lies in the fact that, under SIDM, galaxies can be embedded in haloes with cored central dark matter profiles, irrespective of whether they have a smooth star formation history and non-bursty supernova feedback. In contrast, under CDM, galaxies would require a bursty star formation rate to generate strong supernova feedback that leads the impulsive cusp-core transformation. 

Back to our results, the middle panels of Fig.~\ref{Density_profiles} show that, in $10^{11.5}$ M$_{\odot}$ haloes, baryons impact on the dark matter distribution from the SIDM models. Under CDM, $\rho_{\rm{DM}}(r)$ from the Reference hydrodynamical and DM-only simulations agree, but under SIDM, they diverge. The SIDM DM-only simulations produce lower-density and larger-core profiles compared to the SIDM hydrodynamical simulations, which more closely follow the NFW prediction. The WeakStellarFB model generates cuspier density profiles than the Reference model, both under CDM and SIDM. This result suggests that the increased baryonic concentration in the WeakStellarFB model, relative to the Reference model, enhances the central concentration of the dark matter distribution.

The influence of baryons becomes more pronounced in $10^{12}$ M$_{\odot}$ haloes, as shown in the top panels of Fig.~\ref{Density_profiles}. In all dark matter models (CDM and SIDM), the density profiles between hydrodynamical and DM-only simulations no longer agree. Hydrodynamical-CDM models produce a cuspier $\rho_{\rm{DM}}(r)$ than the NFW profile (in line with predictions from adiabatic contraction models (e.g.~\citealt{Gnedin06}). Similarly, hydrodynamical-SIDM models produce a very cuspy $\rho_{\rm{DM}}(r)$, in contrast to the cored $\rho_{\rm{DM}}$ profiles produced in the DM-only SIDM models. Consistent with our results, previous works by \citet{Elbert18}, \citet{Sameie21} and \citet{Rose22} showed that, under SIDM, dark matter density profiles can be either cuspy or even cuspier than their CDM counterparts, depending on the baryonic concentration. \citet{Sameie21} analysed the density profiles of $10^{12}~\rm{M}_{\odot}$ haloes, modelled in high-resolution zoom-in simulations of SIDM within the FIRE galaxy formation scheme (\citealt{Hopkins18}). Their study showed that SIDM haloes can reach higher and steeper central densities than their CDM counterparts. In a similar approach, \citet{Rose22} presented zoom-in SIDM simulations of Milky Way-like galaxies with the IllustrisTNG (\citealt{Pillepich18}) galaxy formation model. They concluded that baryon contraction begins to have an impact on the density profiles of haloes when their embedded galaxies reach stellar masses of $10^8~\rm{M}_{\odot}$. For higher-mass systems such as groups and clusters, the work of \citet{Robertson21} concluded that the haloes profile strongly depends on the final baryonic distributions. They showed this from the analysis of dark matter halo densities modelled with SIDM and the baryonic physics model of EAGLE (\citealt{Schaye15}) in a zoom-in simulation. Similarly, \citet{Despali19}, employing zoom-in SIDM simulations of galaxies with the IllustrisTNG model, showed that smaller-size galaxies were embedded in cuspy SIDM haloes, while more extended galaxies resided in cored-profile haloes. 

The gravitational influence of baryons not only increases central dark matter densities in SIDM models, but also diversify the haloes' dark matter distribution, which can be seen from the increased scatter around the median density profiles (shaded region in Fig.~\ref{Density_profiles}). This diversity could be attributed to variations in the assembly history of galaxies, influencing whether baryons dominate the central gravitational potential sooner or later. 

In Fig.~\ref{Density_evolution_1}, we investigate how the haloes assembly history shapes the evolution of the DM density profile. We select the 32 most massive haloes (with masses larger than $10^{12}$ M$_{\odot}$) at $z=0$ in the cosmological box from the Reference model under CDM (left panel) and SigmaVel30 (right panel). The panels show the evolution of the median $\rho_{\rm{DM}}$ between redshifts 0 and 2 (coloured lines). The black solid line shows the NFW profile of the haloes at redshift zero (estimated using the concentration-mass relation from \citealt{Correa15c}). The left panel shows that except for the inner few kpc, there is minimal evolution of $\rho_{\rm{DM}}(r)$ under CDM. In this case, haloes formed a cuspy profile by redshift two, the subsequent impact of the central galaxies, through ejection of energy via supernova- and AGN-driven winds, leads to the formation of small cores in the center.

Under SIDM, the haloes' density evolves. At redshift two, the central dark matter density of SIDM haloes is lower than for their CDM counterparts. However, as galaxies in SIDM haloes grow in mass, baryons start to dominate the central potential. In response dark matter particles thermalise through frequent interactions and accumulate in the center of the baryonic-dominated potential. Over time, this results in an overconcentration of dark matter, manifesting as a highly cuspy density profile. This can be seen in the increasing central density of SIDM haloes in the right panel of Fig.~\ref{Density_evolution_1}. In the WeakStellarFB models, however, the situation is slightly different. Due to the early domination of baryons of the central potential, SIDM haloes quickly formed highly cuspy density profiles, with minimal evolution in the redshift range zero to two. Further details are presented in Appendix~\ref{AppendixB}.

In this section we have shown how baryons impact on the dark matter distribution under SIDM and CDM. While not an entirely novel result, this study presents the first cosmological simulations of a galaxy population under different velocity-dependent SIDM models and baryonic feedback schemes. The resulting features of the galaxy population have important implications for studies aiming to constrain SIDM by directly comparing to observational datasets. This is shown and discussed in the next section.

\section{Tully-Fisher Relation}\label{TFRelation_sec}

The galaxy sample from the TangoSIDM simulations is characterized by distinct sizes, varying from highly extended to compact, depending on the stellar feedback model (Reference versus WeakStellarFB,  as illustrated in Fig.~\ref{GalProp_fig}). Simultaneously, the sample includes haloes with distinct dark matter distributions, with SIDM haloes having densities that deviate significantly from the NFW profile, as shown in Fig.~\ref{Density_profiles}. The sample's stellar-to-halo mass relation is consistent with observations (as depicted in the left panels of Fig.~\ref{GalProp_fig}), and therefore haloes of a given mass host galaxies of the correct mass range. In this section, we test our galaxy sample with the stellar-mass Tully-Fisher relation, which establishes a correlation between the stellar mass and circular speed at a characteristic radius of spiral galaxies. First investigated by \citet{Tully1977}, the relation has since become one of the best studied galaxy scaling relations (see e.g. \citealt{Bell01, Ziegler02, Pizagno07,Avila08,Reyes11,Catinella23,Ristea24}), so that numerous studies have delved into its cosmological origin using both semi-analytical approaches and simulations (see e.g. \citealt{Steinmetz99,Dutton12,Cattaneo14,Desmond15,Ferrero16}).

Our analysis in this section demonstrates that when TangoSIDM galaxies are too compact or when dark matter is overly concentrated in the center, their rotation curves peak at much higher velocities than observed. This poses a powerful challenge for the validity of SIDM models. To quantify the significance of this constraint, Section~\ref{MassSize_Sec} assesses which simulated galaxy samples, drawn from the Reference versus WeakStellarFB models under the various dark matter scenarios (presented in Section~\ref{Subsample_sec}), are consistent with the observational sample (introduced in Section~\ref{ObsSample_Sec}). This consistency test implies assessing the likelihood that the two sets of samples (simulated and observational) were drawn from the same, albeit unknown, probability distribution. Following this, in Section~\ref{TFRelation_subsec}, we analyse the deviation of TangoSIDM galaxies from the observed Tully-Fisher relation. Subsequently, we evaluate the statistical significance of this deviation in Section~\ref{Stat_analysis}. 

\begin{figure*} 
\includegraphics[angle=0,width=\textwidth]{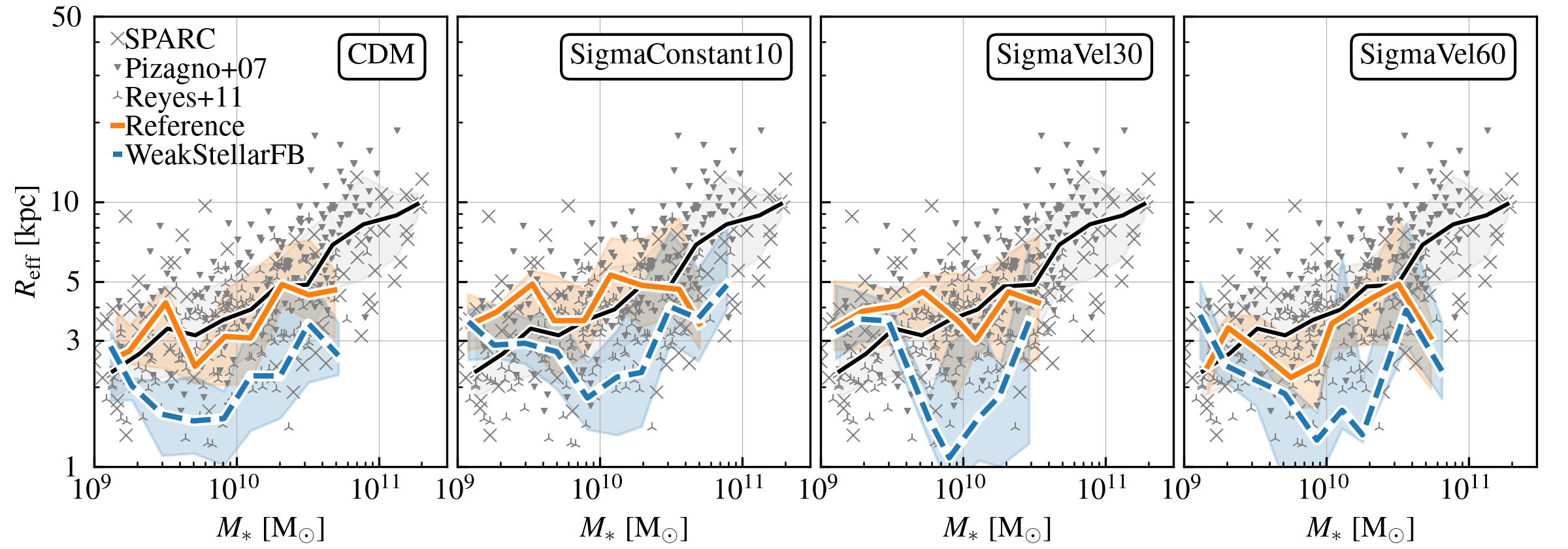}
\caption{Effective radius as a function of stellar mass for $z=0$ disc-type galaxies from the Reference (orange solid line) and WeakStellarFB (blue dashed line) models under CDM (left panel), SigmaConstant10 (second panel from the left), SigmaVel30 (third panel from the left) and SigmaVel60 (right panel). The observational sample is shown in grey symbols, with crosses corresponding to the SPARC dataset, triangles to the \citet{Pizagno07} catalog and stars to the \citet{Reyes11} data. The solid lines indicate the median relations for both the compiled observational sample (black) and the simulated sample (in color), while the shaded regions highlight the 16-84th percentiles. A visual inspection suggests that the simulated samples from the Reference model closely agree with the observational data, whereas the samples from the WeakStellarFB model do not. A statistical analysis using the Kolmogorov-Smirnov test reveals that only the massive ($M_{*}\geq 10^{10}~\rm{M}_{\odot}$) simulated galaxies from the Reference model under CDM, SigmaVel30 and SigmaVel60 are not significantly different from the observational sample.}
\label{Size_plots}
\end{figure*}

\subsection{Simulated sample}\label{Subsample_sec}

We create a subsample of disc-type galaxies using the fraction of stellar kinetic energy invested in ordered co-rotation, $\kappa_{\rm{co}}$, defined as

\begin{equation}\label{kappa}
\kappa_{\rm{co}}=\frac{K_{\rm{co-rot}}}{K}=\frac{1}{K}\sum_{i}^{r<50\rm{kpc}}\frac{1}{2}\,m_i\left[L_{z,i}/(m_i\,R_i)\right]^{2},
\end{equation}

\noindent to quantify morphology (see e.g. \citealt{Correa17}). In eq.~(\ref{kappa}), the sum is over all stellar particles within a spherical radius of 50 kpc centered on the minimum of the potential, $m_{i}$ is the mass of each stellar particle, $K(=\sum_{i}^{r<50\rm{kpc}}\frac{1}{2}m_iv_i^2)$ the total kinetic energy, $L_{z,i}$ the particle angular momentum along the direction of the total angular momentum of the stellar component of the galaxy and $R_{i}$ is the projected distance to the axis of rotation. See also \citet{Sales10} and \citet{Correa20} for more details on $\kappa_{\rm{co}}$.

To create a disc-type galaxy subsample within each simulation, we use the criterion $\kappa_{\rm{co}}{>}0.3$ following \citet{Correa20}, who showed that values in the range $\kappa_{\rm{co}}{=}0.3-0.35$ select disc-type galaxies from the EAGLE simulations that agree with the distribution of disc galaxies from SDSS in the morphology-stellar mass-halo mass plane. This results in a selection of 61 disc-type galaxies per simulation with stellar masses ranging from $10^{9}~\rm{M}_{\odot}$ to $1.2\times 10^{11}~\rm{M}_{\odot}$ and effective sizes, denoted as $R_{\rm{eff}}$, ranging from 1.4 kpc to 17.3 kpc. Note that $R_{\rm{eff}}$ is defined as the 2D projected size enclosing 50 per cent of the total K-band luminosity. The total luminosity is computed from all bound star particles within a projected 2D circular aperture of 50 kpc radius. The luminosities are intrinsic (i.e. dust-free) and are calculated at each output time and for each star particle, accounting for its age, mass, and metallicity. This calculation is performed by the SWIFT code using the photometric tables from \citet{Trayford15}. Finally, we estimate the circular velocity at the effective radius, $V_{\rm{circ}}(R_{\rm{eff}})$, as follows $V_{\rm{circ}}(R_{\rm{eff}})=\sqrt{GM(<R_{\rm{eff}})/R_{\rm{eff}}}$, where the sum $M(<R_{\rm{eff}})$ considers the total mass of  baryons (stars and gas) and dark matter enclosed within $R_{\rm{eff}}$.

\subsection{Observational sample}\label{ObsSample_Sec}

We compile an observational sample by joining the catalogs of disk galaxies from \citet{Lelli16}, \citet{Pizagno07} and \citet{Reyes11}, resulting in a dataset of 429 disc galaxies with stellar masses within the range of $10^{9}~\rm{M}_{\odot}$ to $2\times 10^{11}~\rm{M}_{\odot}$, and effective radii, $R_{\rm{eff}}$, spanning from 1.2 kpc to 18.5 kpc. Note that $R_{\rm{eff}}$ is defined as the radius encompassing half of the total galaxy luminosity. Rotational curves at $R_{\rm{eff}}$ were either directly extracted or estimated from each catalog. In the following, we provide a more detailed overview of these datasets.

\citet{Lelli16} presented the Spitzer Photometry and Accurate Rotation Curves (SPARC) dataset, a galaxy catalog of 175 disc galaxies with near-infrared photometry at 3.6 $\mu$m and well-defined, high-quality HI rotation curves. For our analysis, we extracted inclination-corrected circular velocities, total luminosity at 3.6 $\mu$m, and effective radii directly from SPARC. We followed \citet{Lelli17} and determined stellar masses using a constant mass-to-light ratio of $\Gamma = 0.5~\rm{M}_{\odot}/\rm{L}_{\odot}$, which was motivated by stellar population synthesis models (\citealt{SchombertMcGaugh2014}) using a Chabrier IMF. The total circular velocity at the effective radius was computed by interpolating the rotational curves.

The catalog derived by \citet{Pizagno07} consists of 163 spiral galaxies featuring resolved $H_{\alpha}$ rotation curves. We utilized the effective radius and circular velocity at the effective radius directly from this catalog and estimated stellar masses using the $i$-band magnitudes, assuming a constant I-band mass-to-light ratio of 1.2, $M_{*}=1.2 \times 10^{0.4(i_{\odot}-i)}~\rm{M}_{\odot}$ with $i_{\odot}=4.11$. The mass-to-light-ratio is adopted for a Chabrier IMF and it assumes the contribution of disc+bulge (\citealt{Portinari04}). The effective radius for this sample is defined as the radius at $2.2\times R_{\rm{disk}}$, where $R_{\rm{disk}}$ is the disc exponential scale length.

Finally, \citet{Reyes11} provided an improved estimate of disk rotation velocities for a subset of SDSS galaxies. This dataset includes the $i$-band Petrosian half-light radius, $r$-band Petrosian absolute magnitude ($M_{r}$), and $g-r$ colour, all $k$-corrected to $z = 0$ and corrected for Galactic and internal extinction. Stellar masses were estimated following \citet{Bell03},

\begin{equation}
M_{*}=10^{[\log_{10}(L_{r}/L_{r,\odot})+\log_{10}(M_{*}/L_{r})+\log_{10}h^{2}]}~\rm{M}_{\odot},
\end{equation}

\noindent where $\log_{10}(L_{r}/L_{r,\odot})=-0.4 (M_{r}-M_{r,\odot}+1.1 z)$ with $M_{r,\odot}=4.76$, and $\log_{10}(M_{*}/L_{r})=-0.306+1.097\cdot (g-r)-0.093$, where the last term, $-0.093$, corresponds to the conversion from a modified Salpeter IMF to a Chabrier IMF (as indicated in \citealt{Gallazzi08}).

To estimate the rotation velocity at $R_{\rm{eff}}$, we used the arctangent model 

\begin{equation}
V_{\rm{circ,obs}}(R')=V_{0}+\frac{2}{\pi}V_{\rm{c,obs}}\arctan\left(\frac{R'-R_{0}}{R_{\rm{TO}}}\right).
\end{equation}

\noindent \citet{Reyes11} fitted this model to each rotational curve from the sample and provided the four free parameters: the systemic velocity $V_{0}$, the asymptotic circular velocity $V_{\rm{c,obs}}$, the spatial center $R_{0}$, and the turn-over radius $R_{\rm{TO}}$, at which the rotation curve starts to flatten out. We use the above expression for $V_{\rm{circ,obs}}(R')$ and estimate it at $R_{\rm{eff}}$, by converting $R_{\rm{eff}}$ into arcsecond units and correcting for inclination as follows $V_{\rm{circ}} = V_{\rm{circ,obs}}(R')/\sin(i)$.

\begin{figure*} 
\includegraphics[angle=0,width=\textwidth]{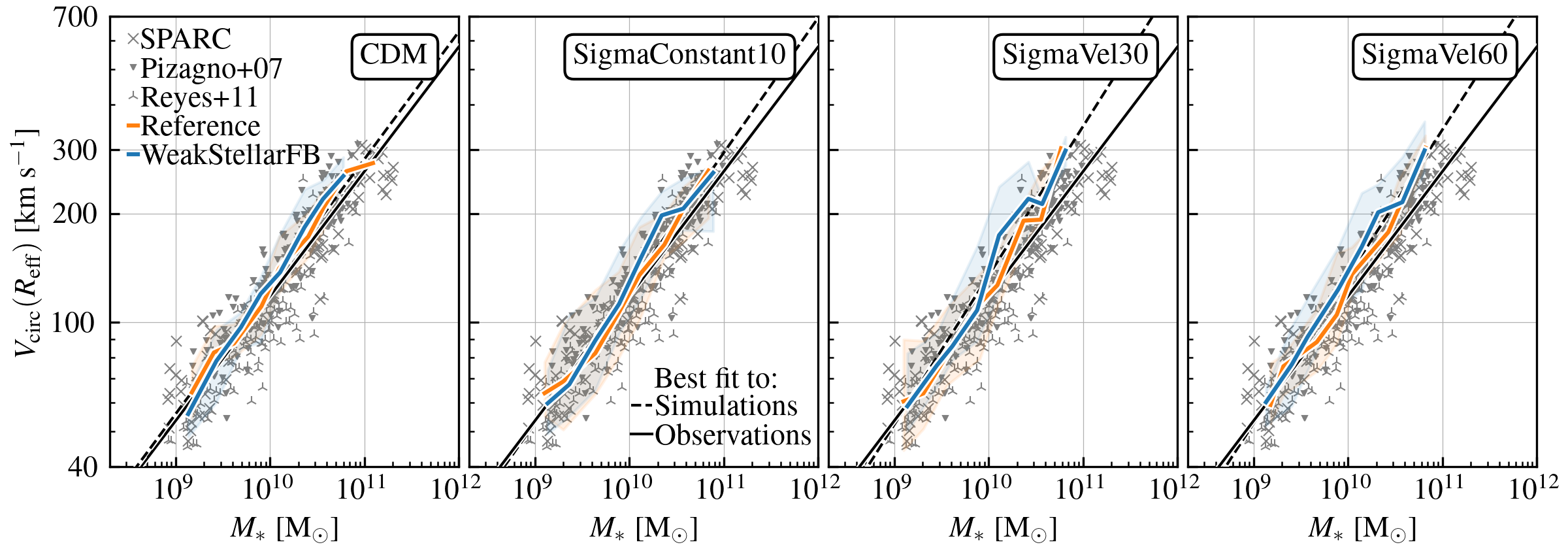}
\caption{The stellar mass Tully-Fisher relation, i.e. total circular velocity at the effective galactic radius as a function of stellar mass for disc-type galaxies. The median relations for disc galaxies in the Reference and WeakStellarFB models are represented by orange and blue lines, respectively, with shaded regions highlighting the $1{-}99$th percentiles. The panels show the Tully-Fisher relation for simulated galaxies under CDM (left panel), SigmaConstant10 (second panel from the left), SigmaVel30 (second panel from the right) and SigmaVel60 (right panel). Similar to Fig.~\ref{Size_plots}, the panels also display the observational sample in grey symbols, with crosses corresponding to the SPARC dataset, triangles to the \citet{Pizagno07} catalog and stars to the \citet{Reyes11} data. The solid and dashed lines depict the best-fitting linear relations to the observational sample and simulated samples, respectively. The figure reveals a close agreement between observations and the Reference and WeakStellarFB models under CDM. However, this agreement is not maintained for SIDM. Under SIDM, the slope of the Tully-Fisher relation from the simulated sample deviates from the observed relation, with the most significant deviation occurring in the SigmaVel30 model, followed by SigmaVel60. The deviation between the relations (observational vs. simulated) is statistically significant at the $98\%$ level in the SigmaVel30 model for galaxies with masses ${\geq}10^{10}~\rm{M}_{\odot}$, and at the $95\%$ confidence level in the SigmaVel60 model for galaxies with masses ${\geq}1.3{\times}10^{10}~\rm{M}_{\odot}$.}
\label{Tully_Fisher_plots}
\end{figure*}

\subsection{The mass-size plane}\label{MassSize_Sec}

We compare the observational sample in the mass-size plane with a sample of disc-type galaxies taken from the simulations. Fig.~\ref{Size_plots} shows the effective radius as a function of stellar mass for disc-type galaxies from the Reference (orange solid line) and WeakStellarFB (blue dashed line) models under CDM (left panel), SigmaConstant10 (second panel from the left), SigmaVel30 (third panel from the left) and SigmaVel60 (right panel). The coloured lines represent the median relations, and the shaded regions depict the 16-84th percentiles. The observational sample is shown in grey symbols, and its median relation is depicted in solid black line.

The panels in Fig.~\ref{Size_plots} indicate that the median trend of the simulated samples from the Reference model agrees with the observational data, while the simulated galaxies from the WeakStellarFB do not, as they become quite compact around a stellar mass of $10^{10}~\rm{M}_{\odot}$. To determine the statistical significance of the differences in the mass-size plane between the simulated and observational samples, we perform a Kolmogorov-Smirnov test (KS) for two samples. Given that the observed sample is not a volume-limited sample, we opt not to account for the mass distribution. Instead, we make a quantitative analysis by dividing the samples into bins of stellar mass and comparing the size distributions. For each stellar mass bin, we test the null hypothesis that the two samples\textemdash observational and simulated\textemdash were drawn from the same distribution. A confidence level of 95\% is chosen, implying that we reject the null hypothesis if the $p$-value is less than 0.05. The aim of the KS test is to identify the simulated galaxy sample that is most likely drawn from the distribution function of the observational sample, making it statistically equivalent.

We separate the samples into three stellar mass bins ([$10^{9}$, $3{\times}10^{9}$], [$3{\times}10^{9}$, $10^{10}$], and [$10^{10}$, $3{\times}10^{10}~\rm{M}_{\odot}$]), and compare the observational sample and simulated galaxies from the Reference model under CDM. The KS statistical analysis returns $p$-values of 0.36, 0.02 and 0.13, respectively under each mass bin. The low $p$-value of 0.02 for simulated galaxies with stellar masses between $3{\times}10^{9}$ and $10^{10}~\rm{M}_{\odot}$ indicates that those galaxies do not conform to the observed size distribution, whereas galaxies in the other mass bins do. We further analyse the samples of galaxies from Reference + SigmaConstant10, SigmaVel30 and SigmaVel60, contrasting them with the observational sample. For Reference + SigmaConstant10, the analysis returns the following $p$-values of $8{\times}10^{-4}$, 0.78 and 0.15. Similarly, Reference + SigmaVel30 returns $p$-values of $8{\times}10^{-4}$, 0.31 and 0.09, whereas Reference + SigmaVel60 yields $p$-values of 0.59, $5{\times}10^{-3}$ and 0.06. For all Reference models (CDM + SIDM), the large $p$-values in the stellar mass bins $10^{10}-3{\times}10^{10}~\rm{M}_{\odot}$ indicate that we cannot reject the null hypothesis. Therefore, at the high mass end the samples, both observational and simulated, are not significantly different at the 95$\%$ confidence level and could be drawn from the same size distribution.

Differently, the WeakStellarFB model (under CDM or SIDM) fails to produce galaxies with sizes that agree with the observations. The KS test returns small $p$-values (${<}0.01$) for galaxies more massive than $3{\times}10^{9}~\rm{M}_{\odot}$. For lower mass galaxies, in the regime where the overcooling of the model has a lesser impact (as discussed in Section~\ref{Gal_prop_section}), the KS test yields $p$-values of 0.06, 0.1 and 0.98 for WeakStellarFB + SigmaConstant10, + SigmaVel30, and + SigmaVel60, respectively. From what we conclude that in the low mass end, the WeakStellarFB model under SIDM, produces galaxies whose sizes are not statistically different from the observations.

\subsection{Tully-Fisher relation}\label{TFRelation_subsec}

The Tully-Fisher relation is shown in Fig.~\ref{Tully_Fisher_plots}, where the y-axis corresponds to the total circular velocity at the effective galactic radius and the x-axis corresponds to the stellar mass. The left panel displays the Tully-Fisher relation for disc galaxies from the Reference (orange solid line) and WeakStellarFB (blue solid line) models under CDM. Moving from left to right, the subsequent panels show the relation for disc galaxies under the SigmaConstant10 model, SigmaVel60 model, and SigmaVel30 model. Similar to Fig.~\ref{Size_plots}, the panels also show the observational sample in grey symbols. Coloured lines highlight the median relations from the simulations, while shaded regions represent the 1-99th percentiles. 

The figure shows a tight correlation between circular velocity and stellar mass, as expected. This correlation is further highlighted by the best-fitting linear relation to the observational sample (black solid lines). The best-fitting parameters of the relation, $\log_{10}(V_{\rm{circ}}/{\rm{km~s^{-1}}})=a\log_{10}(M_{*}/10^{10}{\rm{M}}_{\odot})+b$, are $a=0.34{\pm}0.01$ and $b=2.07{\pm}0.01$. The parameters and $5-95\%$ confidence intervals were estimated by bootstrapping the observational sample.\footnote{In each bootstrap iteration $i$, we created a random observational subsample and estimated the best-fitting parameters $a_{i}$ and $b_{i}$ of the subsample utilizing the {\texttt{stats.linregress}} function from the {\texttt{scipy}} package.} Similarly, we created a joint sample of galaxies from both the Reference and WeakStellarFB models, and via the bootstrap method we estimated the best-fitting linear relations from the simulations, which are depicted by black dashed lines in the panels.

Fig.~\ref{Tully_Fisher_plots} shows the close agreement between the Tully-Fisher relation derived from the observational sample and that of the simulated sample of disc galaxies from both the Reference and WeakStellarFB models under CDM. However, this agreement is not maintained when considering the SIDM models. Under the SIDM framework, the slope of the Tully-Fisher relation from the simulated sample begins to deviate relative to the observed Tully-Fisher relation. The largest deviation occurs in the SigmaVel30 model, followed by the SigmaVel60 model. The shift in $V_{\rm{circ}}(R_{\rm{eff}})$ found in galaxies within the SIDM models is attributed to the large central dark matter densities that result from the dark matter particle interactions. Consequently, at constant $M_{*}$, the increased enclosed dark matter mass drives a higher $V_{\rm{circ}}(R_{\rm{eff}})$ compared to CDM, thereby altering the slope of the relation.

Our analysis in the previous subsection established that only the disc galaxies from the Reference model under CDM, SigmaVel30 and SigmaVel60 were statistically comparable to the observational sample. This was not found for galaxies from the WeakStellarFB model under any dark matter model. Nonetheless, in Fig.~\ref{Tully_Fisher_plots}, we intentionally include the trend from the WeakStellarFB model to highlight how the deviation from the observed Tully-Fisher relation increases under SIDM, particularly when galaxies become more compact. Notably, under CDM, the Tully-Fisher relations from both the Reference and WeakStellarFB models closely agree. This  finding appears to contradict the conclusions drawn by \citet{Ferrero16}, who posited that $\Lambda$CDM models should be capable of matching the observed Tully-Fisher relation, provided that galaxy sizes are well reproduced, and that halos respond approximately adiabatically to galaxy assembly. This will be further addressed in future work, with more variations of the stellar feedback model and larger number statistics from the simulated sample.

\subsection{Statistical analysis}\label{Stat_analysis}

The panels in Fig.~\ref{Tully_Fisher_plots} reveal a discernible departure of the Tully-Fisher relation from disc galaxies under SIDM relative to the observed Tully-Fisher relation. To quantify the significance of this deviation and to assess the likelihood of a similar deviation in the observational sample, we perform a statistical analysis focusing on the observational sample and the simulated samples from the Reference model under SigmaVel30 and SigmaVel60. Reference+SigmaConstant10 is not considered in the analysis because this particular SIDM model has already been ruled out by observations of galaxy clusters. Additionally, the WeakStellarFB models are excluded from the analysis due to their significant difference from the observations (as established in Section~\ref{MassSize_Sec}).

Given that the deviation in the SIDM models is prominent in massive galaxies, as demonstrated in Section~\ref{Density_sec} for haloes more massive than $10^{12}~\rm{M}_{\odot}$, and considering that these galaxies are statistically equivalent to the observational sample in terms of their size distribution, as demonstrated in Section~\ref{MassSize_Sec}, we apply a selection cut in stellar mass of $10^{10}~\rm{M}_{\odot}$. This allows us to analyse the Tully-Fisher relation for only massive galaxies, while disregarding the influence of low-mass systems that do not present significant changes in their central density profiles relative to CDM. This stellar mass cut results in a subsample of 287 real galaxies and 33 simulated galaxies from each SIDM model. 

We perform a bootstrap analysis around these subsamples using 10,000 iterations. In each iteration, we create random samples (with replacement) for both the observational and simulated datasets, and calculate the slopes of their respective Tully-Fisher relations. After all iterations, we calculate the mean value and confidence intervals of the slopes. For the observational sample, a slope of $a=0.34{\pm}0.02$ is obtained. When comparing this slope with its value for the entire sample (calculated in Section~\ref{TFRelation_subsec}), we find that the observed Tully-Fisher relation does not change when we consider only the subsample of massive galaxies. 

For the simulated sample from the SigmaVel30 model, we find a slope of $a=0.48{\pm}0.08$, and for the SigmaVel60 model, a slope of $a=0.41{\pm}0.07$. The SigmaVel30 model exhibits a strong deviation from the observed Tully-Fisher relation, as indicated by the different slope. When we assess the differences between these slopes, we obtain a $p$-value of 0.012, which indicates the frequency that each random bootstrap sample from the simulations had a slope lower than the slope from the observational bootstrap sample. Consequently, we conclude that the Reference+SigmaVel30 model, despite producing galaxies with stellar masses and sizes in good agreement with the observations (Fig.~\ref{Size_plots}), produces a Tully-Fisher relation that deviates from the observed one at the $98\%$ confidence level.

The Tully-Fisher relation from the SigmaVel60 model also deviates from the observed relation, although it is not as pronounced as in the SigmaVel30 case. The difference between these slopes yields a $p$-value of 0.13, signifying that in ${\sim}13\%$ of the bootstrap samples from the SIDM model, a slope equivalent or lower than the one derived from the observations arises. Therefore, we cannot reject the null hypothesis that both Tully-Fisher relations, from the observations and simulations, are drawn from the same distribution. 

We further investigate this and calculate the minimum stellar mass above which the simulated galaxy sample from SigmaVel60 produces a Tully-Fisher relation that deviates significantly from the observed one. This cut is identified for galaxies with $M_{*}\geq 1.3{\times}10^{10}~\rm{M}_{\odot}$. For these refined subsamples, the observational sample yields a slope of $a=0.32{\pm}0.03$, while the SigmaVel60 model produces a slope of  $a=0.45{\pm}0.09$. The bootstrap analysis yields a low $p$-value of 0.042, indicating that over this mass range, the SigmaVel60 model produces a Tully-Fisher relation that deviates from the observed one at the $95\%$ confidence level. As a control test, we assess the difference in the Tully-Fisher relations from the Reference+CDM model and observations over this mass range of ${>}1.3\times 10^{10}~\rm{M}_{\odot}$. We obtain a $p$-value of 0.26. Thus, we affirm that the Reference+CDM model maintains a good agreement with the observations.

The deviation of the Tully-Fisher relation from disc galaxies (under the Reference+SigmaVel30 models) relative to the observed Tully-Fisher relation is non-negligible. In this section, we have shown that it is statistically significant, which indicates that we can rule out the SigmaVel30 model over the mass range ${\gtrsim}10^{10}~\rm{M}_{\odot}$ with $98\%$ confidence. The rejection of this SIDM model relies on the assumption that the Reference galaxy formation model is valid, as we have demonstrated through the good agreement of the stellar mass-halo mass relation (Fig.~\ref{GalProp_fig}), stellar masses (Fig.~\ref{SMF_fig}) and galaxy sizes (Fig.~\ref{Size_plots}, supported by statistical analysis of Section~\ref{MassSize_Sec}) with observations. 

The deviation of the Tully-Fisher relation, relative to observations, is driven by the impact of SIDM, which produces haloes with high central dark matter densities (Fig.~\ref{Density_profiles}). SIDM therefore raises $V_{\rm{circ}}(R_{\rm{eff}})$ at fixed stellar mass, and increases the slope of the relation as shown in this section. This physical effect, constrained by the Tully-Fisher relation, is ruled out. Note, however, that the rejection of the SigmaVel30 model is specific to a certain ``mass range" (i.e. ${\gtrsim}10^{10}~\rm{M}_{\odot}$), because halo density evolution depends on the value of the cross section, which in turn depends on halo mass (Fig.~\ref{sidm_models}). Therefore, only cross sections influencing the steepness of the haloes' density throughout their evolution are ruled out. In a similar manner, we argue that the Tully-Fisher relation can be utilized to rule out the SigmaVel60 model over the mass range ${\gtrsim}1.3\times 10^{10}~\rm{M}_{\odot}$ with $95\%$ confidence. These findings have significant implications for the SIDM parameter space, which are discussed in the next section.

\begin{figure} 
\includegraphics[angle=0,width=0.5\textwidth]{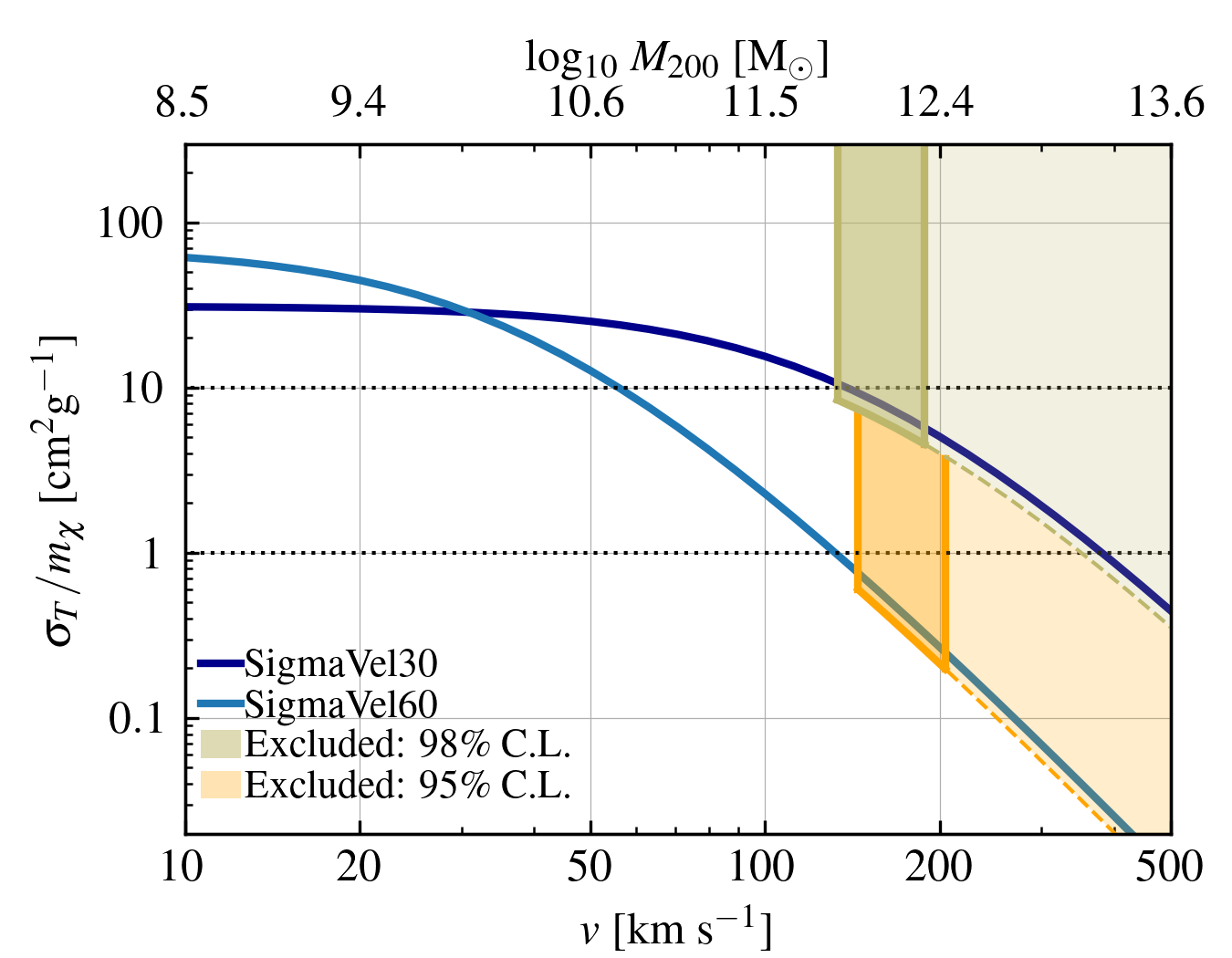}
\caption{Momentum transfer cross section, $\sigma_{\rm{T}}/m_{\chi}$, plotted as a function of relative scattering velocity of dark matter particles. The blue solid lines shows the velocity-depenent SIDM models presented in this work, SigmaVel30 and SigmaVel60 (see Table~\ref{Table_sidm_models} for details). The bottom x-axis indicates the relative velocity between dark matter particles, while the top x-axis indicates the typical halo mass that hosts orbits of such velocities. The shaded regions demarcate areas of the SIDM plane excluded by this work. The dark green and orange shaded regions highlight the excluded parameter space that is directly extracted from the simulations. The lighter green and orange shaded regions mark larger regions that are excluded based on the assumption that higher mass haloes under SIDM models with larger cross sections would exhibit a large deviation in the Tully-Fisher plane from the observations.}
\label{SIDM_plot}
\end{figure}

\section{Discussion}\label{Discussion_sec}

\subsection{SIDM parameter space}\label{SIDM_param_space}

The SIDM parameter space, characterized by the self-interaction cross section as a function of the relative velocities between dark matter particles, is a topic of extensive debate. Robust constraints on the cross section on large scales (high dark matter particle velocities) have been established by studies of galaxy clusters (e.g. \citealt{Randall08,Dawson13,Massey15,Harvey15,Wittman18,Harvey19,Sagunski21,Andrade22}). However, the cross section for Milky Way-size galaxies and lower-mass systems remains highly uncertain. Recent proposals suggest that the cross section in dwarf-size galaxies should be as large as $100~\rm{cm}^{2}\rm{g}^{-1}$ (e.g. \citealt{Correa21,Turner21,Silverman23,Slone23,Yang23}), in order to address the diversity problem through halo core expansion and core collapse. At the scale of Milky Way-mass galaxies, \citet{Correa23b} argues that the cross section should be lower than $10~\rm{cm}^{2}\rm{g}^{-1}$. Otherwise the frequent interactions between the Milky Way-mass systems and their satellites would lead to excessive mass loss and destruction of satellites, giving rise to unrealistic satellite populations. 

This section discusses the new constraints on the SIDM parameter space presented in Section~\ref{Stat_analysis}. Our work has shown that the co-evolution of baryons and dark matter self-interactions strongly impacts the evolution of galaxies. Compared with CDM, galaxies in SIDM hydrodynamical simulations not only tend to grow more extended (Section~\ref{Gal_prop_section}), but also contain enhanced dark matter central densities (Section~\ref{Density_sec}). This behavior results in a deviation in the Tully-Fisher relation relative to an observational dataset (Section~\ref{TFRelation_subsec}). In Section~\ref{Stat_analysis}, we found that this deviation is statistically significant in the SigmaVel30 model for galaxies more massive than ${\gtrsim}10^{10}~\rm{M}_{\odot}$, and in the SigmaVel60 model for ${\gtrsim}1.3\times 10^{10}~\rm{M}_{\odot}$. Next, we place these constraints on the velocity-$\sigma_{\rm{T}}/m_{\chi}$ plane. 

In what follows we argue that we can rule out velocity-cross section pairs that govern the evolution of haloes hosting the massive disc galaxies that significantly deviate from the observations in the Tully-Fisher plane. To identify the velocity-cross section pairs, we therefore select all disc galaxies from the Reference + SigmaVel30 and Reference + SigmaVel60 models with stellar masses larger than $10^{10}~\rm{M}_{\odot}$ and $1.3{\times}10^{10}~\rm{M}_{\odot}$, respectively. We follow the assembly histories of the haloes hosting these galaxies across the simulation snapshots until redshift 2, the redshift below which the haloes' density profiles are well resolved and commence substantial evolution (refer to Fig.~\ref{Density_evolution_1}, Section~\ref{Density_sec} and Appendix~\ref{AppendixB}). We determine the median and 16-84th percentiles of their mass accretion histories, $M_{200}(z)$, and convert these to the circular velocity, $V_{\rm{circ}}(z)$. We assume that $V_{\rm{circ}}(z)$ is the average velocity of the dark matter particles within these haloes over the redshift range 0-2, and using eqs.~(\ref{sigmam}) and (\ref{sigmat}) we estimate the haloes' average dark matter cross section. The individual haloes' $M_{200}(z)$, $V_{\rm{circ}}(z)$ and cross sections are shown in Appendix~\ref{velocity_sigma_pairs}.

The derived velocity-cross section pairs establish the limits above which the SigmaVel30 and SigmaVel60 models produce overly enhanced central dark matter densities in massive disc galaxies. Therefore, we mark these limits as regions where the SigmaVel30 and SigmaVel60 models are ruled out with $98\%$ and $95\%$ confidence, as shown in the green and orange shaded areas in Fig.~\ref{SIDM_plot}. The figure depicts the momentum transfer cross section, $\sigma_{\rm{T}}/m_{\chi}$, as a function of relative dark matter particle scattering velocity. The curves show the velocity-dependent SIDM models presented in this work, SigmaVel30 and SigmaVel60 (see Table~\ref{Table_sidm_models}). While the bottom x-axis highlights the relative velocity between dark matter particles, the top x-axis indicates the typical halo mass that hosts orbits of such velocities. The dark green and orange shaded regions highlight the newly excluded parameter space that is directly extracted from the simulations. The lighter green and orange shaded regions mark further excluded regions under the assumption that higher mass haloes under SIDM models with higher cross sections would exhibit a large deviation in the Tully-Fisher plane from the observations.

While this finding imposes strong constraints on velocity-dependent models, it does not entirely rule them out. There is still room for models where the cross section reaches $100~\rm{cm}^{2}\rm{g}^{-1}$ at 10 km s$^{-1}$, provided that it decreases to less than $1~\rm{cm}^{2}\rm{g}^{-1}$ at 150 km s$^{-1}$. In Fig.~\ref{SIDM_plot}, we refrain from extending the SIDM parameter space to velocities larger than 500 km s$^{-1}$, since those are not covered by the simulations. Our future plans include expanding this analysis to larger scales, employing larger cosmological boxes and more statistical power through increased numerical resolution to model a more extensive sample with lower mass disc galaxies. Additionally, we aim to explore the circular velocities of dwarf galaxies in more detail. We anticipate that with sufficient resolution and statistics, the modelling of dwarf galaxies, even with the inclusion of baryons as shown in the bottom panels of Fig.~\ref{Density_profiles}, may yield lower values of $V_{\rm{circ}}(R_{\rm{eff}})$ relative to an observational sample at fixed stellar mass, consequently resulting in a deviation of the Tully-Fisher relation. This analysis, coupled with methodology improvements such as mock observations of HI discs for extracting rotational curves, will be the focus of future work.

\section{Conclusions}\label{Conclusion_sec}

The SIDM parameter space, while extensively explored in recent years, remains notably uncertain, particular for Milky Way-size galaxies and smaller systems. This uncertainty arises due to the inherent challenge of isolating the impact of baryonic physics from dark matter interactions. Recent studies (e.g.~\citealt{Robertson19,Despali19,Sameie21,Rose22,Burger22,Jiang23}) have reported that the prevalence of baryons in the central gravitational potential leads to the formation of denser and more cusp-like central density profiles under SIDM compared to CDM. Nevertheless, uncertainties persist regarding how the increased cuspiness of SIDM haloes correlates with the specific SIDM model parameters and the strength of galaxy feedback. To address these uncertainties, this study introduces a new set of cosmological hydrodynamical simulations. These simulations include the SIDM model derived from the TangoSIDM project (\citealt{Correa22}) and leverage the baryonic physics from the SWIFT-EAGLE galaxy formation model (\citealt{Borrow22b,Schaller23}). 

Two cosmological simulation suites were generated: The Reference model, calibrated in a (25 Mpc)$^3$ volume to reproduce the galaxy stellar mass function and galaxy mass-size relation; and the WeakStellarFB model, featuring less efficient stellar feedback around Milky Way-like systems. Each galaxy formation model (Reference and WeakStellarFB) was simulated under four dark matter cosmologies: CDM, SigmaConstant10 (a SIDM model with a constant cross section of 10 cm$^{2}\rm{g}^{-1}$), and SigmaVel30 and SigmaVel60, two SIDM models with velocity-dependent cross sections (see Fig.~\ref{sidm_models}). SigmaVel60 has a cross section smaller than 1 cm$^{2}\rm{g}^{-1}$ at high velocities ($v{>}150$ km s$^{-1}$) and increases with decreasing velocity, reaching 60 cm$^{2}\rm{g}^{-1}$ at 10 km s$^{-1}$. SigmaVel30 has a cross section smaller than 8 cm$^{2}\rm{g}^{-1}$ at velocities surpassing $200$ km s$^{-1}$ (dropping below 1 cm$^{2}\rm{g}^{-1}$ at ${\approx}1000$ km s$^{-1}$) and it also increases with decreasing velocity. These SIDM models we selected to represent two extreme scenarios for the rate of dark matter interactions in Milky Way-mass systems. The SWIFT-EAGLE models were selected to determine whether the impact of SIDM on galaxies remains robust when subjected to variations in feedback models. 

Our findings indicate that SIDM does not significantly alter global galaxy properties such as stellar masses and star formation rates, but it does impact galaxy sizes, making galaxies more extended (Fig.~\ref{GalProp_fig}). Dark matter particle interactions heat the inner halo, leading to core formation in the central regions of haloes less massive than $10^{11}~\rm{M}_{\odot}$ and dynamically heating the surrounding gas and stars, promoting the formation of more extended galaxies. However, we have found that the impact of SIDM is insufficient to counteract the gas overcooling and size compactness in galaxies from the WeakStellarFB model. 

In massive haloes (${\sim}10^{12}~\rm{M}_{\odot}$), baryonic influence on SIDM distributions result in steeper dark matter density profiles than those produced in CDM from adiabatic contraction (Fig.~\ref{Density_profiles}). This feature is enhanced in the WeakStellarFB model, suggesting that the increased baryonic concentration in the model, relative to the Reference model, enhances the central concentration of the dark matter distribution. Under SIDM, the haloes density profile evolved differently (Fig.~\ref{Density_evolution_1}). As galaxies grow in mass, baryons begin to dominate the central gravitational potential, causing dark matter particles to thermalise through frequent interactions and accumulate in the center, resulting in cuspy dark matter density profiles.

The enhanced dark matter density at the centers of galaxies results in a notable deviation in the slope of the Tully-Fisher relation, significantly diverging from observations. We assembled an observational sample of $z\approx 0$ disc galaxies by combining the catalogs from \citet{Pizagno07}, \citet{Reyes11} and \citet{Lelli16}. Our analysis reveals that while the simulated massive galaxies from the Reference model under SigmaVel30 and SigmaVel60 are not significantly different from the observational sample in the galaxy mass-size plane (Fig.~\ref{Size_plots}), they strongly deviate in the Tully-Fisher plane (Fig.~\ref{Tully_Fisher_plots}). This is due to a shift in $V_{\rm{circ}}(R_{\rm{eff}})$ found in galaxies within the SIDM models, driven by the large central dark matter densities that result from the dark matter particle interactions. Consequently, at constant $M_{*}$, the increased enclosed dark matter mass leads to a higher $V_{\rm{circ}}(R_{\rm{eff}})$ compared to CDM, altering the slope of the relation. In contrast, the Tully-Fisher relation derived from CDM models aligns well with observations. 

We have conducted a statistical analysis to assess the significance of the discrepancy between the SIDM models and observations in the Tully-Fisher plane. Our findings indicate that galaxies from the Reference+SigmaVel30 model more massive than $10^{10}~\rm{M}_{\odot}$ deviate from the observational sample at the 98\% confidence level, while galaxies with masses exceeding $1.3{\times}10^{10}~\rm{M}_{\odot}$ from the Reference+SigmaVel60 model deviate at the 95\% confidence level. These constraints, when translated into the velocity-$\sigma_{\rm{T}}/m_{\chi}$ plane (Fig.~\ref{SIDM_plot}), reveal that the cross section should be smaller than 0.5 cm$^{2}\rm{g}^{-1}$ for velocities of $\sim$150-200 km s$^{-1}$ and smaller than 10 cm$^{2}\rm{g}^{-1}$ for velocities of 110-180 km s$^{-1}$.

Our study reveals that the Tully-Fisher plane, encompassing galaxy sizes, stellar masses, and circular velocities, serves as a powerful observable for discerning and excluding velocity-dependent SIDM models. In future work we will focus on improving the datasets (higher numerical resolution and larger cosmological box size for the simulations, as well as larger data compilation from observational surveys) and refining the methodology, including the creation of mock HI rotation curves, with the goal of carrying out more accurate and precise comparisons.

\section*{Acknowledgements}

CC acknowledges the support of the Dutch Research Council (NWO Veni 192.020). YMB gratefully acknowledges funding from the Netherlands Organization for Scientific Research (NWO) under Veni grant number 639.041.751 and financial support from the Swiss National Science Foundation (SNSF) under funding reference 200021/\ 213076. The research in this paper made use of the SWIFT open-source simulation code (http://www.swiftsim.com, \citealt{Schaller23}). The TangoSIDM simulation suite have been produced using the DECI resource Mahti based in Finland at CSC, Finnish IT Center for Science, with support from the PRACE aisbl., project ID 17DECI0030-TangoSIDM. The TangoSIDM simulations design and analysis has been carried using the Dutch national e-infrastructure, Snellius, with the support of SURF Cooperative, project ID EINF-180-TangoSIDM. We acknowledge various public python packages that have greatly benefited this work: \verb|scipy| (\citealt{vanderWalt11}), \verb|numpy| (\citealt{vanderWalt11}), \verb|matplotlib| (\citealt{Hunter07}). This work has also benefited from the python analysis pipeline SwiftsimIO (\citealt{Borrow20,Borrow21}). 

\section*{Data availability}

The data supporting the plots within this article will be made publicly available in http://www.tangosidm.com. See example in Table~\ref{Table_observational_data}.

\bibliography{biblio}

\begin{thebibliography}{}
\makeatletter
\relax
\def\mn@urlcharsother{\let\do\@makeother \do\$\do\&\do\#\do\^\do\_\do\%\do\~}
\def\mn@doi{\begingroup\mn@urlcharsother \@ifnextchar [ {\mn@doi@}
  {\mn@doi@[]}}
\def\mn@doi@[#1]#2{\def\@tempa{#1}\ifx\@tempa\@empty \href
  {http://dx.doi.org/#2} {doi:#2}\else \href {http://dx.doi.org/#2} {#1}\fi
  \endgroup}
\def\mn@eprint#1#2{\mn@eprint@#1:#2::\@nil}
\def\mn@eprint@arXiv#1{\href {http://arxiv.org/abs/#1} {{\tt arXiv:#1}}}
\def\mn@eprint@dblp#1{\href {http://dblp.uni-trier.de/rec/bibtex/#1.xml}
  {dblp:#1}}
\def\mn@eprint@#1:#2:#3:#4\@nil{\def\@tempa {#1}\def\@tempb {#2}\def\@tempc
  {#3}\ifx \@tempc \@empty \let \@tempc \@tempb \let \@tempb \@tempa \fi \ifx
  \@tempb \@empty \def\@tempb {arXiv}\fi \@ifundefined
  {mn@eprint@\@tempb}{\@tempb:\@tempc}{\expandafter \expandafter \csname
  mn@eprint@\@tempb\endcsname \expandafter{\@tempc}}}

\bibitem[\protect\citeauthoryear{{Adhikari} et~al.,}{{Adhikari}
  et~al.}{2022}]{Adhikari22}
{Adhikari} S.,  et~al., 2022, \mn@doi [arXiv e-prints]
  {10.48550/arXiv.2207.10638}, \href
  {https://ui.adsabs.harvard.edu/abs/2022arXiv220710638A} {p. arXiv:2207.10638}

\bibitem[\protect\citeauthoryear{{Andrade}, {Fuson}, {Gad-Nasr}, {Kong},
  {Minor}, {Roberts}  \& {Kaplinghat}}{{Andrade} et~al.}{2022}]{Andrade22}
{Andrade} K.~E.,  {Fuson} J.,  {Gad-Nasr} S.,  {Kong} D.,  {Minor} Q.,
  {Roberts} M.~G.,   {Kaplinghat} M.,  2022, \mn@doi [\mnras]
  {10.1093/mnras/stab3241}, \href
  {https://ui.adsabs.harvard.edu/abs/2022MNRAS.510...54A} {510, 54}

\bibitem[\protect\citeauthoryear{{Arkani-Hamed}, {Finkbeiner}, {Slatyer}  \&
  {Weiner}}{{Arkani-Hamed} et~al.}{2009}]{ArkaniHamed09}
{Arkani-Hamed} N.,  {Finkbeiner} D.~P.,  {Slatyer} T.~R.,   {Weiner} N.,  2009,
  \mn@doi [\prd] {10.1103/PhysRevD.79.015014}, \href
  {https://ui.adsabs.harvard.edu/abs/2009PhRvD..79a5014A} {79, 015014}

\bibitem[\protect\citeauthoryear{{Avila-Reese}, {Zavala}, {Firmani}  \&
  {Hern{\'a}ndez-Toledo}}{{Avila-Reese} et~al.}{2008}]{Avila08}
{Avila-Reese} V.,  {Zavala} J.,  {Firmani} C.,   {Hern{\'a}ndez-Toledo} H.~M.,
  2008, \mn@doi [\aj] {10.1088/0004-6256/136/3/1340}, \href
  {https://ui.adsabs.harvard.edu/abs/2008AJ....136.1340A} {136, 1340}

\bibitem[\protect\citeauthoryear{{Bah{\'e}} et~al.,}{{Bah{\'e}}
  et~al.}{2022}]{Bahe22}
{Bah{\'e}} Y.~M.,  et~al., 2022, \mn@doi [\mnras] {10.1093/mnras/stac1339},
  \href {https://ui.adsabs.harvard.edu/abs/2022MNRAS.516..167B} {516, 167}

\bibitem[\protect\citeauthoryear{{Banerjee}, {Adhikari}, {Dalal}, {More}  \&
  {Kravtsov}}{{Banerjee} et~al.}{2020}]{Banerjee20}
{Banerjee} A.,  {Adhikari} S.,  {Dalal} N.,  {More} S.,   {Kravtsov} A.,  2020,
  \mn@doi [\jcap] {10.1088/1475-7516/2020/02/024}, \href
  {https://ui.adsabs.harvard.edu/abs/2020JCAP...02..024B} {2020, 024}

\bibitem[\protect\citeauthoryear{{Bauer} et~al.,}{{Bauer}
  et~al.}{2013}]{Bauer13}
{Bauer} A.~E.,  et~al., 2013, \mn@doi [\mnras] {10.1093/mnras/stt1011}, \href
  {https://ui.adsabs.harvard.edu/abs/2013MNRAS.434..209B} {434, 209}

\bibitem[\protect\citeauthoryear{{Behroozi}, {Wechsler}, {Hearin}  \&
  {Conroy}}{{Behroozi} et~al.}{2019}]{Behroozi19}
{Behroozi} P.,  {Wechsler} R.~H.,  {Hearin} A.~P.,   {Conroy} C.,  2019,
  \mn@doi [\mnras] {10.1093/mnras/stz1182}, \href
  {https://ui.adsabs.harvard.edu/abs/2019MNRAS.488.3143B} {488, 3143}

\bibitem[\protect\citeauthoryear{{Bell} \& {de Jong}}{{Bell} \& {de
  Jong}}{2001}]{Bell01}
{Bell} E.~F.,  {de Jong} R.~S.,  2001, \mn@doi [\apj] {10.1086/319728}, \href
  {https://ui.adsabs.harvard.edu/abs/2001ApJ...550..212B} {550, 212}

\bibitem[\protect\citeauthoryear{{Bell}, {McIntosh}, {Katz}  \&
  {Weinberg}}{{Bell} et~al.}{2003}]{Bell03}
{Bell} E.~F.,  {McIntosh} D.~H.,  {Katz} N.,   {Weinberg} M.~D.,  2003, \mn@doi
  [\apjs] {10.1086/378847}, \href
  {https://ui.adsabs.harvard.edu/abs/2003ApJS..149..289B} {149, 289}

\bibitem[\protect\citeauthoryear{{Boddy}, {Feng}, {Kaplinghat}, {Shadmi}  \&
  {Tait}}{{Boddy} et~al.}{2014}]{Boddy14}
{Boddy} K.~K.,  {Feng} J.~L.,  {Kaplinghat} M.,  {Shadmi} Y.,   {Tait} T.
  M.~P.,  2014, \mn@doi [\prd] {10.1103/PhysRevD.90.095016}, \href
  {https://ui.adsabs.harvard.edu/abs/2014PhRvD..90i5016B} {90, 095016}

\bibitem[\protect\citeauthoryear{{Booth} \& {Schaye}}{{Booth} \&
  {Schaye}}{2009}]{Booth09}
{Booth} C.~M.,  {Schaye} J.,  2009, \mn@doi [\mnras]
  {10.1111/j.1365-2966.2009.15043.x}, \href
  {https://ui.adsabs.harvard.edu/abs/2009MNRAS.398...53B} {398, 53}

\bibitem[\protect\citeauthoryear{{Borrow} \& {Borrisov}}{{Borrow} \&
  {Borrisov}}{2020}]{Borrow20}
{Borrow} J.,  {Borrisov} A.,  2020, \mn@doi [The Journal of Open Source
  Software] {10.21105/joss.02430}, \href
  {https://ui.adsabs.harvard.edu/abs/2020JOSS....5.2430B} {5, 2430}

\bibitem[\protect\citeauthoryear{{Borrow} \& {Kelly}}{{Borrow} \&
  {Kelly}}{2021}]{Borrow21}
{Borrow} J.,  {Kelly} A.~J.,  2021, arXiv e-prints, \href
  {https://ui.adsabs.harvard.edu/abs/2021arXiv210605281B} {p. arXiv:2106.05281}

\bibitem[\protect\citeauthoryear{{Borrow}, {Schaller}, {Bower}  \&
  {Schaye}}{{Borrow} et~al.}{2022}]{Borrow22}
{Borrow} J.,  {Schaller} M.,  {Bower} R.~G.,   {Schaye} J.,  2022, \mn@doi
  [\mnras] {10.1093/mnras/stab3166}, \href
  {https://ui.adsabs.harvard.edu/abs/2022MNRAS.511.2367B} {511, 2367}

\bibitem[\protect\citeauthoryear{{Borrow}, {Schaller}, {Bah{\'e}}, {Schaye},
  {Ludlow}, {Ploeckinger}, {Nobels}  \& {Altamura}}{{Borrow}
  et~al.}{2023}]{Borrow22b}
{Borrow} J.,  {Schaller} M.,  {Bah{\'e}} Y.~M.,  {Schaye} J.,  {Ludlow} A.~D.,
  {Ploeckinger} S.,  {Nobels} F. S.~J.,   {Altamura} E.,  2023, \mn@doi
  [\mnras] {10.1093/mnras/stad2928}, \href
  {https://ui.adsabs.harvard.edu/abs/2023MNRAS.526.2441B} {526, 2441}

\bibitem[\protect\citeauthoryear{{Borukhovetskaya}, {Navarro}, {Errani}  \&
  {Fattahi}}{{Borukhovetskaya} et~al.}{2022}]{Borukhovetskaya22}
{Borukhovetskaya} A.,  {Navarro} J.~F.,  {Errani} R.,   {Fattahi} A.,  2022,
  \mn@doi [\mnras] {10.1093/mnras/stac653}, \href
  {https://ui.adsabs.harvard.edu/abs/2022MNRAS.512.5247B} {512, 5247}

\bibitem[\protect\citeauthoryear{{Brook}, {Stinson}, {Gibson}, {Ro{\v{s}}kar},
  {Wadsley}  \& {Quinn}}{{Brook} et~al.}{2012}]{Brook12}
{Brook} C.~B.,  {Stinson} G.,  {Gibson} B.~K.,  {Ro{\v{s}}kar} R.,  {Wadsley}
  J.,   {Quinn} T.,  2012, \mn@doi [\mnras] {10.1111/j.1365-2966.2011.19740.x},
  \href {https://ui.adsabs.harvard.edu/abs/2012MNRAS.419..771B} {419, 771}

\bibitem[\protect\citeauthoryear{{Buckley} \& {Fox}}{{Buckley} \&
  {Fox}}{2010}]{Buckley10}
{Buckley} M.~R.,  {Fox} P.~J.,  2010, \mn@doi [\prd]
  {10.1103/PhysRevD.81.083522}, \href
  {https://ui.adsabs.harvard.edu/abs/2010PhRvD..81h3522B} {81, 083522}

\bibitem[\protect\citeauthoryear{{Burger}, {Zavala}, {Sales}, {Vogelsberger},
  {Marinacci}  \& {Torrey}}{{Burger} et~al.}{2022}]{Burger22}
{Burger} J.~D.,  {Zavala} J.,  {Sales} L.~V.,  {Vogelsberger} M.,  {Marinacci}
  F.,   {Torrey} P.,  2022, \mn@doi [\mnras] {10.1093/mnras/stac994}, \href
  {https://ui.adsabs.harvard.edu/abs/2022MNRAS.513.3458B} {513, 3458}

\bibitem[\protect\citeauthoryear{{Ca{\~n}as}, {Elahi}, {Welker}, {del P Lagos},
  {Power}, {Dubois}  \& {Pichon}}{{Ca{\~n}as} et~al.}{2019}]{Canas19}
{Ca{\~n}as} R.,  {Elahi} P.~J.,  {Welker} C.,  {del P Lagos} C.,  {Power} C.,
  {Dubois} Y.,   {Pichon} C.,  2019, \mn@doi [\mnras] {10.1093/mnras/sty2725},
  \href {https://ui.adsabs.harvard.edu/abs/2019MNRAS.482.2039C} {482, 2039}

\bibitem[\protect\citeauthoryear{{Catinella} et~al.,}{{Catinella}
  et~al.}{2023}]{Catinella23}
{Catinella} B.,  et~al., 2023, \mn@doi [\mnras] {10.1093/mnras/stac3556}, \href
  {https://ui.adsabs.harvard.edu/abs/2023MNRAS.519.1098C} {519, 1098}

\bibitem[\protect\citeauthoryear{{Cattaneo}, {Salucci}  \&
  {Papastergis}}{{Cattaneo} et~al.}{2014}]{Cattaneo14}
{Cattaneo} A.,  {Salucci} P.,   {Papastergis} E.,  2014, \mn@doi [\apj]
  {10.1088/0004-637X/783/2/66}, \href
  {https://ui.adsabs.harvard.edu/abs/2014ApJ...783...66C} {783, 66}

\bibitem[\protect\citeauthoryear{{Chabrier}}{{Chabrier}}{2003}]{Chabrier03}
{Chabrier} G.,  2003, \mn@doi [\pasp] {10.1086/376392}, \href
  {https://ui.adsabs.harvard.edu/abs/2003PASP..115..763C} {115, 763}

\bibitem[\protect\citeauthoryear{{Chaikin}, {Schaye}, {Schaller}, {Bah{\'e}},
  {Nobels}  \& {Ploeckinger}}{{Chaikin} et~al.}{2022}]{Chaikin22}
{Chaikin} E.,  {Schaye} J.,  {Schaller} M.,  {Bah{\'e}} Y.~M.,  {Nobels} F.
  S.~J.,   {Ploeckinger} S.,  2022, \mn@doi [\mnras] {10.1093/mnras/stac1132},
  \href {https://ui.adsabs.harvard.edu/abs/2022MNRAS.514..249C} {514, 249}

\bibitem[\protect\citeauthoryear{{Chang}, {van der Wel}, {da Cunha}  \&
  {Rix}}{{Chang} et~al.}{2015}]{Chang15}
{Chang} Y.-Y.,  {van der Wel} A.,  {da Cunha} E.,   {Rix} H.-W.,  2015, \mn@doi
  [\apjs] {10.1088/0067-0049/219/1/8}, \href
  {https://ui.adsabs.harvard.edu/abs/2015ApJS..219....8C} {219, 8}

\bibitem[\protect\citeauthoryear{{Col{\'\i}n}, {Avila-Reese}, {Valenzuela}  \&
  {Firmani}}{{Col{\'\i}n} et~al.}{2002}]{Colin02}
{Col{\'\i}n} P.,  {Avila-Reese} V.,  {Valenzuela} O.,   {Firmani} C.,  2002,
  \mn@doi [\apj] {10.1086/344259}, \href
  {https://ui.adsabs.harvard.edu/abs/2002ApJ...581..777C} {581, 777}

\bibitem[\protect\citeauthoryear{{Correa}}{{Correa}}{2021}]{Correa21}
{Correa} C.~A.,  2021, \mn@doi [\mnras] {10.1093/mnras/stab506}, \href
  {https://ui.adsabs.harvard.edu/abs/2021MNRAS.503..920C} {503, 920}

\bibitem[\protect\citeauthoryear{Correa}{Correa}{2023}]{Correa23b}
Correa C.~A.,  2023, \mn@doi [SciPost Phys. Proc.]
  {10.21468/SciPostPhysProc.12.059}, p.~059

\bibitem[\protect\citeauthoryear{{Correa} \& {Schaye}}{{Correa} \&
  {Schaye}}{2020}]{Correa20}
{Correa} C.~A.,  {Schaye} J.,  2020, \mn@doi [\mnras] {10.1093/mnras/staa3053},
  \href {https://ui.adsabs.harvard.edu/abs/2020MNRAS.499.3578C} {499, 3578}

\bibitem[\protect\citeauthoryear{{Correa}, {Wyithe}, {Schaye}  \&
  {Duffy}}{{Correa} et~al.}{2015}]{Correa15c}
{Correa} C.~A.,  {Wyithe} J. S.~B.,  {Schaye} J.,   {Duffy} A.~R.,  2015,
  \mn@doi [\mnras] {10.1093/mnras/stv697}, \href
  {https://ui.adsabs.harvard.edu/abs/2015MNRAS.450.1521C} {450, 1521}

\bibitem[\protect\citeauthoryear{{Correa}, {Schaye}, {Clauwens}, {Bower},
  {Crain}, {Schaller}, {Theuns}  \& {Thob}}{{Correa} et~al.}{2017}]{Correa17}
{Correa} C.~A.,  {Schaye} J.,  {Clauwens} B.,  {Bower} R.~G.,  {Crain} R.~A.,
  {Schaller} M.,  {Theuns} T.,   {Thob} A. C.~R.,  2017, \mn@doi [\mnras]
  {10.1093/mnrasl/slx133}, \href
  {https://ui.adsabs.harvard.edu/abs/2017MNRAS.472L..45C} {472, L45}

\bibitem[\protect\citeauthoryear{{Correa}, {Schaller}, {Ploeckinger}, {Anau
  Montel}, {Weniger}  \& {Ando}}{{Correa} et~al.}{2022}]{Correa22}
{Correa} C.~A.,  {Schaller} M.,  {Ploeckinger} S.,  {Anau Montel} N.,
  {Weniger} C.,   {Ando} S.,  2022, \mn@doi [\mnras] {10.1093/mnras/stac2830},
  \href {https://ui.adsabs.harvard.edu/abs/2022MNRAS.517.3045C} {517, 3045}

\bibitem[\protect\citeauthoryear{{Crain} et~al.,}{{Crain}
  et~al.}{2015}]{Crain15}
{Crain} R.~A.,  et~al., 2015, \mn@doi [\mnras] {10.1093/mnras/stv725}, \href
  {https://ui.adsabs.harvard.edu/abs/2015MNRAS.450.1937C} {450, 1937}

\bibitem[\protect\citeauthoryear{{Dalla Vecchia} \& {Schaye}}{{Dalla Vecchia}
  \& {Schaye}}{2012}]{DallaVecchia12}
{Dalla Vecchia} C.,  {Schaye} J.,  2012, \mn@doi [\mnras]
  {10.1111/j.1365-2966.2012.21704.x}, \href
  {https://ui.adsabs.harvard.edu/abs/2012MNRAS.426..140D} {426, 140}

\bibitem[\protect\citeauthoryear{{Dav{\'e}}, {Spergel}, {Steinhardt}  \&
  {Wandelt}}{{Dav{\'e}} et~al.}{2001}]{Dave01}
{Dav{\'e}} R.,  {Spergel} D.~N.,  {Steinhardt} P.~J.,   {Wandelt} B.~D.,  2001,
  \mn@doi [\apj] {10.1086/318417}, \href
  {https://ui.adsabs.harvard.edu/abs/2001ApJ...547..574D} {547, 574}

\bibitem[\protect\citeauthoryear{{Dawson} et~al.,}{{Dawson}
  et~al.}{2013}]{Dawson13}
{Dawson} W.,  et~al., 2013, in American Astronomical Society Meeting Abstracts
  \#221. p. 125.04

\bibitem[\protect\citeauthoryear{{Desmond} \& {Wechsler}}{{Desmond} \&
  {Wechsler}}{2015}]{Desmond15}
{Desmond} H.,  {Wechsler} R.~H.,  2015, \mn@doi [\mnras]
  {10.1093/mnras/stv1978}, \href
  {https://ui.adsabs.harvard.edu/abs/2015MNRAS.454..322D} {454, 322}

\bibitem[\protect\citeauthoryear{{Despali}, {Sparre}, {Vegetti},
  {Vogelsberger}, {Zavala}  \& {Marinacci}}{{Despali} et~al.}{2019}]{Despali19}
{Despali} G.,  {Sparre} M.,  {Vegetti} S.,  {Vogelsberger} M.,  {Zavala} J.,
  {Marinacci} F.,  2019, \mn@doi [\mnras] {10.1093/mnras/stz273}, \href
  {https://ui.adsabs.harvard.edu/abs/2019MNRAS.484.4563D} {484, 4563}

\bibitem[\protect\citeauthoryear{{Dooley}, {Peter}, {Vogelsberger}, {Zavala}
  \& {Frebel}}{{Dooley} et~al.}{2016}]{Dooley16}
{Dooley} G.~A.,  {Peter} A. H.~G.,  {Vogelsberger} M.,  {Zavala} J.,   {Frebel}
  A.,  2016, \mn@doi [\mnras] {10.1093/mnras/stw1309}, \href
  {https://ui.adsabs.harvard.edu/abs/2016MNRAS.461..710D} {461, 710}

\bibitem[\protect\citeauthoryear{{Driver} et~al.,}{{Driver}
  et~al.}{2022}]{Driver22}
{Driver} S.~P.,  et~al., 2022, \mn@doi [\mnras] {10.1093/mnras/stac472}, \href
  {https://ui.adsabs.harvard.edu/abs/2022MNRAS.513..439D} {513, 439}

\bibitem[\protect\citeauthoryear{{Dutton} \& {van den Bosch}}{{Dutton} \& {van
  den Bosch}}{2012}]{Dutton12}
{Dutton} A.~A.,  {van den Bosch} F.~C.,  2012, \mn@doi [\mnras]
  {10.1111/j.1365-2966.2011.20339.x}, \href
  {https://ui.adsabs.harvard.edu/abs/2012MNRAS.421..608D} {421, 608}

\bibitem[\protect\citeauthoryear{{Elahi}, {Thacker}  \& {Widrow}}{{Elahi}
  et~al.}{2011}]{Elahi11}
{Elahi} P.~J.,  {Thacker} R.~J.,   {Widrow} L.~M.,  2011, \mn@doi [\mnras]
  {10.1111/j.1365-2966.2011.19485.x}, \href
  {https://ui.adsabs.harvard.edu/abs/2011MNRAS.418..320E} {418, 320}

\bibitem[\protect\citeauthoryear{{Elahi}, {Ca{\~n}as}, {Poulton}, {Tobar},
  {Willis}, {Lagos}, {Power}  \& {Robotham}}{{Elahi} et~al.}{2019}]{Elahi19}
{Elahi} P.~J.,  {Ca{\~n}as} R.,  {Poulton} R. J.~J.,  {Tobar} R.~J.,  {Willis}
  J.~S.,  {Lagos} C. d.~P.,  {Power} C.,   {Robotham} A. S.~G.,  2019, \mn@doi
  [\pasa] {10.1017/pasa.2019.12}, \href
  {https://ui.adsabs.harvard.edu/abs/2019PASA...36...21E} {36, e021}

\bibitem[\protect\citeauthoryear{{Elbert}, {Bullock}, {Kaplinghat},
  {Garrison-Kimmel}, {Graus}  \& {Rocha}}{{Elbert} et~al.}{2018}]{Elbert18}
{Elbert} O.~D.,  {Bullock} J.~S.,  {Kaplinghat} M.,  {Garrison-Kimmel} S.,
  {Graus} A.~S.,   {Rocha} M.,  2018, \mn@doi [\apj]
  {10.3847/1538-4357/aa9710}, \href
  {https://ui.adsabs.harvard.edu/abs/2018ApJ...853..109E} {853, 109}

\bibitem[\protect\citeauthoryear{{Faucher-Gigu{\`e}re}}{{Faucher-Gigu{\`e}re}}{2020}]{FaucherGiguere20}
{Faucher-Gigu{\`e}re} C.-A.,  2020, \mn@doi [\mnras] {10.1093/mnras/staa302},
  \href {https://ui.adsabs.harvard.edu/abs/2020MNRAS.493.1614F} {493, 1614}

\bibitem[\protect\citeauthoryear{{Feng}, {Kaplinghat}  \& {Yu}}{{Feng}
  et~al.}{2010}]{Feng10}
{Feng} J.~L.,  {Kaplinghat} M.,   {Yu} H.-B.,  2010, \mn@doi [\prd]
  {10.1103/PhysRevD.82.083525}, \href
  {https://ui.adsabs.harvard.edu/abs/2010PhRvD..82h3525F} {82, 083525}

\bibitem[\protect\citeauthoryear{{Ferrero} et~al.,}{{Ferrero}
  et~al.}{2017}]{Ferrero16}
{Ferrero} I.,  et~al., 2017, \mn@doi [\mnras] {10.1093/mnras/stw2691}, \href
  {https://ui.adsabs.harvard.edu/abs/2017MNRAS.464.4736F} {464, 4736}

\bibitem[\protect\citeauthoryear{{Gallazzi}, {Brinchmann}, {Charlot}  \&
  {White}}{{Gallazzi} et~al.}{2008}]{Gallazzi08}
{Gallazzi} A.,  {Brinchmann} J.,  {Charlot} S.,   {White} S. D.~M.,  2008,
  \mn@doi [\mnras] {10.1111/j.1365-2966.2007.12632.x}, \href
  {https://ui.adsabs.harvard.edu/abs/2008MNRAS.383.1439G} {383, 1439}

\bibitem[\protect\citeauthoryear{{Gilman}, {Bovy}, {Treu}, {Nierenberg},
  {Birrer}, {Benson}  \& {Sameie}}{{Gilman} et~al.}{2021}]{Gilman21}
{Gilman} D.,  {Bovy} J.,  {Treu} T.,  {Nierenberg} A.,  {Birrer} S.,  {Benson}
  A.,   {Sameie} O.,  2021, \mn@doi [\mnras] {10.1093/mnras/stab2335}, \href
  {https://ui.adsabs.harvard.edu/abs/2021MNRAS.507.2432G} {507, 2432}

\bibitem[\protect\citeauthoryear{{Gilman}, {Zhong}  \& {Bovy}}{{Gilman}
  et~al.}{2023}]{Gilman23}
{Gilman} D.,  {Zhong} Y.-M.,   {Bovy} J.,  2023, \mn@doi [\prd]
  {10.1103/PhysRevD.107.103008}, \href
  {https://ui.adsabs.harvard.edu/abs/2023PhRvD.107j3008G} {107, 103008}

\bibitem[\protect\citeauthoryear{{Gnedin}}{{Gnedin}}{2006}]{Gnedin06}
{Gnedin} O.~Y.,  2006, in {Mamon} G.~A.,  {Combes} F.,  {Deffayet} C.,   {Fort}
  B.,  eds,  EAS Publications Series Vol. 20, EAS Publications Series. pp
  59--64 (\mn@eprint {arXiv} {astro-ph/0510539}), \mn@doi{10.1051/eas:2006048}

\bibitem[\protect\citeauthoryear{{Greengard} \& {Rokhlin}}{{Greengard} \&
  {Rokhlin}}{1987}]{Greengard87}
{Greengard} L.,  {Rokhlin} V.,  1987, \mn@doi [Journal of Computational
  Physics] {10.1016/0021-9991(87)90140-9}, \href
  {https://ui.adsabs.harvard.edu/abs/1987JCoPh..73..325G} {73, 325}

\bibitem[\protect\citeauthoryear{{Harvey}, {Massey}, {Kitching}, {Taylor}  \&
  {Tittley}}{{Harvey} et~al.}{2015}]{Harvey15}
{Harvey} D.,  {Massey} R.,  {Kitching} T.,  {Taylor} A.,   {Tittley} E.,  2015,
  \mn@doi [Science] {10.1126/science.1261381}, \href
  {https://ui.adsabs.harvard.edu/abs/2015Sci...347.1462H} {347, 1462}

\bibitem[\protect\citeauthoryear{{Harvey}, {Robertson}, {Massey}  \&
  {McCarthy}}{{Harvey} et~al.}{2019}]{Harvey19}
{Harvey} D.,  {Robertson} A.,  {Massey} R.,   {McCarthy} I.~G.,  2019, \mn@doi
  [\mnras] {10.1093/mnras/stz1816}, \href
  {https://ui.adsabs.harvard.edu/abs/2019MNRAS.488.1572H} {488, 1572}

\bibitem[\protect\citeauthoryear{{Hayashi}, {Ibe}, {Kobayashi}, {Nakayama}  \&
  {Shirai}}{{Hayashi} et~al.}{2021}]{Hayashi21}
{Hayashi} K.,  {Ibe} M.,  {Kobayashi} S.,  {Nakayama} Y.,   {Shirai} S.,  2021,
  \mn@doi [\prd] {10.1103/PhysRevD.103.023017}, \href
  {https://ui.adsabs.harvard.edu/abs/2021PhRvD.103b3017H} {103, 023017}

\bibitem[\protect\citeauthoryear{{Hopkins} et~al.,}{{Hopkins}
  et~al.}{2018}]{Hopkins18}
{Hopkins} P.~F.,  et~al., 2018, \mn@doi [\mnras] {10.1093/mnras/sty1690}, \href
  {https://ui.adsabs.harvard.edu/abs/2018MNRAS.480..800H} {480, 800}

\bibitem[\protect\citeauthoryear{Hunter}{Hunter}{2007}]{Hunter07}
Hunter J.~D.,  2007, \mn@doi [Computing in Science \& Engineering]
  {10.1109/MCSE.2007.55}, 9, 90

\bibitem[\protect\citeauthoryear{{Ibe} \& {Yu}}{{Ibe} \& {Yu}}{2010}]{Ibe10}
{Ibe} M.,  {Yu} H.-B.,  2010, \mn@doi [Physics Letters B]
  {10.1016/j.physletb.2010.07.026}, \href
  {https://ui.adsabs.harvard.edu/abs/2010PhLB..692...70I} {692, 70}

\bibitem[\protect\citeauthoryear{{Jenkins}}{{Jenkins}}{2010}]{Jenkins10}
{Jenkins} A.,  2010, \mn@doi [\mnras] {10.1111/j.1365-2966.2010.16259.x}, \href
  {https://ui.adsabs.harvard.edu/abs/2010MNRAS.403.1859J} {403, 1859}

\bibitem[\protect\citeauthoryear{{Jenkins}}{{Jenkins}}{2013}]{Jenkins13}
{Jenkins} A.,  2013, \mn@doi [\mnras] {10.1093/mnras/stt1154}, \href
  {https://ui.adsabs.harvard.edu/abs/2013MNRAS.434.2094J} {434, 2094}

\bibitem[\protect\citeauthoryear{{Jiang} et~al.,}{{Jiang}
  et~al.}{2023}]{Jiang23}
{Jiang} F.,  et~al., 2023, \mn@doi [\mnras] {10.1093/mnras/stad705}, \href
  {https://ui.adsabs.harvard.edu/abs/2023MNRAS.521.4630J} {521, 4630}

\bibitem[\protect\citeauthoryear{{Kahlhoefer}, {Schmidt-Hoberg}, {Kummer}  \&
  {Sarkar}}{{Kahlhoefer} et~al.}{2015}]{Kahlhoefer15}
{Kahlhoefer} F.,  {Schmidt-Hoberg} K.,  {Kummer} J.,   {Sarkar} S.,  2015,
  \mn@doi [\mnras] {10.1093/mnrasl/slv088}, \href
  {https://ui.adsabs.harvard.edu/abs/2015MNRAS.452L..54K} {452, L54}

\bibitem[\protect\citeauthoryear{{Kahlhoefer}, {Kaplinghat}, {Slatyer}  \&
  {Wu}}{{Kahlhoefer} et~al.}{2019}]{Kahlhoefer19}
{Kahlhoefer} F.,  {Kaplinghat} M.,  {Slatyer} T.~R.,   {Wu} C.-L.,  2019,
  \mn@doi [\jcap] {10.1088/1475-7516/2019/12/010}, \href
  {https://ui.adsabs.harvard.edu/abs/2019JCAP...12..010K} {2019, 010}

\bibitem[\protect\citeauthoryear{{Kennicutt}}{{Kennicutt}}{1998}]{Kennicutt98}
{Kennicutt} Robert~C. J.,  1998, \mn@doi [\apj] {10.1086/305588}, \href
  {https://ui.adsabs.harvard.edu/abs/1998ApJ...498..541K} {498, 541}

\bibitem[\protect\citeauthoryear{{Kugel} \& {Borrow}}{{Kugel} \&
  {Borrow}}{2022}]{Kugel22}
{Kugel} R.,  {Borrow} J.,  2022, \mn@doi [The Journal of Open Source Software]
  {10.21105/joss.04240}, \href
  {https://ui.adsabs.harvard.edu/abs/2022JOSS....7.4240K} {7, 4240}

\bibitem[\protect\citeauthoryear{{Kummer}, {Br{\"u}ggen}, {Dolag}, {Kahlhoefer}
   \& {Schmidt-Hoberg}}{{Kummer} et~al.}{2019}]{Kummer19}
{Kummer} J.,  {Br{\"u}ggen} M.,  {Dolag} K.,  {Kahlhoefer} F.,
  {Schmidt-Hoberg} K.,  2019, \mn@doi [\mnras] {10.1093/mnras/stz1261}, \href
  {https://ui.adsabs.harvard.edu/abs/2019MNRAS.487..354K} {487, 354}

\bibitem[\protect\citeauthoryear{{Lange} et~al.,}{{Lange}
  et~al.}{2015}]{Lange15}
{Lange} R.,  et~al., 2015, \mn@doi [\mnras] {10.1093/mnras/stu2467}, \href
  {https://ui.adsabs.harvard.edu/abs/2015MNRAS.447.2603L} {447, 2603}

\bibitem[\protect\citeauthoryear{{Lelli}, {McGaugh}  \& {Schombert}}{{Lelli}
  et~al.}{2016}]{Lelli16}
{Lelli} F.,  {McGaugh} S.~S.,   {Schombert} J.~M.,  2016, \mn@doi [\aj]
  {10.3847/0004-6256/152/6/157}, \href
  {https://ui.adsabs.harvard.edu/abs/2016AJ....152..157L} {152, 157}

\bibitem[\protect\citeauthoryear{{Lelli}, {McGaugh}, {Schombert}  \&
  {Pawlowski}}{{Lelli} et~al.}{2017}]{Lelli17}
{Lelli} F.,  {McGaugh} S.~S.,  {Schombert} J.~M.,   {Pawlowski} M.~S.,  2017,
  \mn@doi [\apj] {10.3847/1538-4357/836/2/152}, \href
  {https://ui.adsabs.harvard.edu/abs/2017ApJ...836..152L} {836, 152}

\bibitem[\protect\citeauthoryear{{Ludlow}, {Schaye}, {Schaller}  \&
  {Richings}}{{Ludlow} et~al.}{2019a}]{Ludlow19}
{Ludlow} A.~D.,  {Schaye} J.,  {Schaller} M.,   {Richings} J.,  2019a, \mn@doi
  [\mnras] {10.1093/mnrasl/slz110}, \href
  {https://ui.adsabs.harvard.edu/abs/2019MNRAS.488L.123L} {488, L123}

\bibitem[\protect\citeauthoryear{{Ludlow}, {Schaye}  \& {Bower}}{{Ludlow}
  et~al.}{2019b}]{Ludlow19b}
{Ludlow} A.~D.,  {Schaye} J.,   {Bower} R.,  2019b, \mn@doi [\mnras]
  {10.1093/mnras/stz1821}, \href
  {https://ui.adsabs.harvard.edu/abs/2019MNRAS.488.3663L} {488, 3663}

\bibitem[\protect\citeauthoryear{{Ludlow}, {Fall}, {Wilkinson}, {Schaye}  \&
  {Obreschkow}}{{Ludlow} et~al.}{2023}]{Ludlow23}
{Ludlow} A.~D.,  {Fall} S.~M.,  {Wilkinson} M.~J.,  {Schaye} J.,   {Obreschkow}
  D.,  2023, \mn@doi [\mnras] {10.1093/mnras/stad2615}, \href
  {https://ui.adsabs.harvard.edu/abs/2023MNRAS.525.5614L} {525, 5614}

\bibitem[\protect\citeauthoryear{{Massey} et~al.,}{{Massey}
  et~al.}{2015}]{Massey15}
{Massey} R.,  et~al., 2015, \mn@doi [\mnras] {10.1093/mnras/stv467}, \href
  {https://ui.adsabs.harvard.edu/abs/2015MNRAS.449.3393M} {449, 3393}

\bibitem[\protect\citeauthoryear{{McAlpine} et~al.,}{{McAlpine}
  et~al.}{2016}]{McAlpine16}
{McAlpine} S.,  et~al., 2016, \mn@doi [Astronomy and Computing]
  {10.1016/j.ascom.2016.02.004}, \href
  {https://ui.adsabs.harvard.edu/abs/2016A&C....15...72M} {15, 72}

\bibitem[\protect\citeauthoryear{{Nadler}, {Banerjee}, {Adhikari}, {Mao}  \&
  {Wechsler}}{{Nadler} et~al.}{2020}]{Nadler20}
{Nadler} E.~O.,  {Banerjee} A.,  {Adhikari} S.,  {Mao} Y.-Y.,   {Wechsler}
  R.~H.,  2020, arXiv e-prints, \href
  {https://ui.adsabs.harvard.edu/abs/2020arXiv200108754N} {p. arXiv:2001.08754}

\bibitem[\protect\citeauthoryear{{Nadler}, {Yang}  \& {Yu}}{{Nadler}
  et~al.}{2023}]{Nadler23}
{Nadler} E.~O.,  {Yang} D.,   {Yu} H.-B.,  2023, \mn@doi [\apjl]
  {10.3847/2041-8213/ad0e09}, \href
  {https://ui.adsabs.harvard.edu/abs/2023ApJ...958L..39N} {958, L39}

\bibitem[\protect\citeauthoryear{{Oman} et~al.,}{{Oman} et~al.}{2015}]{Oman15}
{Oman} K.~A.,  et~al., 2015, \mn@doi [\mnras] {10.1093/mnras/stv1504}, \href
  {https://ui.adsabs.harvard.edu/abs/2015MNRAS.452.3650O} {452, 3650}

\bibitem[\protect\citeauthoryear{{Peter}, {Rocha}, {Bullock}  \&
  {Kaplinghat}}{{Peter} et~al.}{2013}]{Peter13}
{Peter} A. H.~G.,  {Rocha} M.,  {Bullock} J.~S.,   {Kaplinghat} M.,  2013,
  \mn@doi [\mnras] {10.1093/mnras/sts535}, \href
  {https://ui.adsabs.harvard.edu/abs/2013MNRAS.430..105P} {430, 105}

\bibitem[\protect\citeauthoryear{{Pillepich} et~al.,}{{Pillepich}
  et~al.}{2018}]{Pillepich18}
{Pillepich} A.,  et~al., 2018, \mn@doi [\mnras] {10.1093/mnras/stx2656}, \href
  {https://ui.adsabs.harvard.edu/abs/2018MNRAS.473.4077P} {473, 4077}

\bibitem[\protect\citeauthoryear{{Pizagno} et~al.,}{{Pizagno}
  et~al.}{2007}]{Pizagno07}
{Pizagno} J.,  et~al., 2007, \mn@doi [\aj] {10.1086/519522}, \href
  {https://ui.adsabs.harvard.edu/abs/2007AJ....134..945P} {134, 945}

\bibitem[\protect\citeauthoryear{{Planck Collaboration} et~al.}{{Planck
  Collaboration} et~al.}{2014}]{Planck13}
{Planck Collaboration} et~al., 2014, \mn@doi [\aap]
  {10.1051/0004-6361/201321591}, \href
  {http://adsabs.harvard.edu/abs/2014A%26A...571A..16P} {571, A16}

\bibitem[\protect\citeauthoryear{{Ploeckinger} \& {Schaye}}{{Ploeckinger} \&
  {Schaye}}{2020}]{Ploeckinger20}
{Ploeckinger} S.,  {Schaye} J.,  2020, \mn@doi [\mnras]
  {10.1093/mnras/staa2172}, \href
  {https://ui.adsabs.harvard.edu/abs/2020MNRAS.497.4857P} {497, 4857}

\bibitem[\protect\citeauthoryear{{Portinari}, {Sommer-Larsen}  \&
  {Tantalo}}{{Portinari} et~al.}{2004}]{Portinari04}
{Portinari} L.,  {Sommer-Larsen} J.,   {Tantalo} R.,  2004, \mn@doi [\mnras]
  {10.1111/j.1365-2966.2004.07207.x}, \href
  {https://ui.adsabs.harvard.edu/abs/2004MNRAS.347..691P} {347, 691}

\bibitem[\protect\citeauthoryear{{Pospelov}, {Ritz}  \& {Voloshin}}{{Pospelov}
  et~al.}{2008}]{Pospelov08}
{Pospelov} M.,  {Ritz} A.,   {Voloshin} M.,  2008, \mn@doi [\prd]
  {10.1103/PhysRevD.78.115012}, \href
  {https://ui.adsabs.harvard.edu/abs/2008PhRvD..78k5012P} {78, 115012}

\bibitem[\protect\citeauthoryear{{Power}, {Navarro}, {Jenkins}, {Frenk},
  {White}, {Springel}, {Stadel}  \& {Quinn}}{{Power} et~al.}{2003}]{Power03}
{Power} C.,  {Navarro} J.~F.,  {Jenkins} A.,  {Frenk} C.~S.,  {White} S.~D.~M.,
   {Springel} V.,  {Stadel} J.,   {Quinn} T.,  2003, \mn@doi [\mnras]
  {10.1046/j.1365-8711.2003.05925.x}, \href
  {https://ui.adsabs.harvard.edu/abs/2003MNRAS.338...14P} {338, 14}

\bibitem[\protect\citeauthoryear{{Randall}, {Markevitch}, {Clowe}, {Gonzalez}
  \& {Brada{\v{c}}}}{{Randall} et~al.}{2008}]{Randall08}
{Randall} S.~W.,  {Markevitch} M.,  {Clowe} D.,  {Gonzalez} A.~H.,
  {Brada{\v{c}}} M.,  2008, \mn@doi [\apj] {10.1086/587859}, \href
  {https://ui.adsabs.harvard.edu/abs/2008ApJ...679.1173R} {679, 1173}

\bibitem[\protect\citeauthoryear{{Reyes}, {Mandelbaum}, {Gunn}, {Pizagno}  \&
  {Lackner}}{{Reyes} et~al.}{2011}]{Reyes11}
{Reyes} R.,  {Mandelbaum} R.,  {Gunn} J.~E.,  {Pizagno} J.,   {Lackner} C.~N.,
  2011, \mn@doi [\mnras] {10.1111/j.1365-2966.2011.19415.x}, \href
  {https://ui.adsabs.harvard.edu/abs/2011MNRAS.417.2347R} {417, 2347}

\bibitem[\protect\citeauthoryear{{Ristea}, {Cortese}, {Fraser-McKelvie},
  {Catinella}, {van de Sande}, {Croom}  \& {Swinbank}}{{Ristea}
  et~al.}{2024}]{Ristea24}
{Ristea} A.,  {Cortese} L.,  {Fraser-McKelvie} A.,  {Catinella} B.,  {van de
  Sande} J.,  {Croom} S.~M.,   {Swinbank} A.~M.,  2024, \mn@doi [\mnras]
  {10.1093/mnras/stad3638}, \href
  {https://ui.adsabs.harvard.edu/abs/2024MNRAS.527.7438R} {527, 7438}

\bibitem[\protect\citeauthoryear{{Robertson}, {Massey}  \& {Eke}}{{Robertson}
  et~al.}{2017}]{Robertson17}
{Robertson} A.,  {Massey} R.,   {Eke} V.,  2017, \mn@doi [\mnras]
  {10.1093/mnras/stx463}, \href
  {https://ui.adsabs.harvard.edu/abs/2017MNRAS.467.4719R} {467, 4719}

\bibitem[\protect\citeauthoryear{{Robertson}, {Harvey}, {Massey}, {Eke},
  {McCarthy}, {Jauzac}, {Li}  \& {Schaye}}{{Robertson}
  et~al.}{2019}]{Robertson19}
{Robertson} A.,  {Harvey} D.,  {Massey} R.,  {Eke} V.,  {McCarthy} I.~G.,
  {Jauzac} M.,  {Li} B.,   {Schaye} J.,  2019, \mn@doi [\mnras]
  {10.1093/mnras/stz1815}, \href
  {https://ui.adsabs.harvard.edu/abs/2019MNRAS.488.3646R} {488, 3646}

\bibitem[\protect\citeauthoryear{{Robertson}, {Massey}, {Eke}, {Schaye}  \&
  {Theuns}}{{Robertson} et~al.}{2021}]{Robertson21}
{Robertson} A.,  {Massey} R.,  {Eke} V.,  {Schaye} J.,   {Theuns} T.,  2021,
  \mn@doi [\mnras] {10.1093/mnras/staa3954}, \href
  {https://ui.adsabs.harvard.edu/abs/2021MNRAS.501.4610R} {501, 4610}

\bibitem[\protect\citeauthoryear{{Robles} et~al.,}{{Robles}
  et~al.}{2017}]{Robles17}
{Robles} V.~H.,  et~al., 2017, \mn@doi [\mnras] {10.1093/mnras/stx2253}, \href
  {https://ui.adsabs.harvard.edu/abs/2017MNRAS.472.2945R} {472, 2945}

\bibitem[\protect\citeauthoryear{{Robles}, {Kelley}, {Bullock}  \&
  {Kaplinghat}}{{Robles} et~al.}{2019}]{Robles19}
{Robles} V.~H.,  {Kelley} T.,  {Bullock} J.~S.,   {Kaplinghat} M.,  2019,
  \mn@doi [\mnras] {10.1093/mnras/stz2345}, \href
  {https://ui.adsabs.harvard.edu/abs/2019MNRAS.490.2117R} {490, 2117}

\bibitem[\protect\citeauthoryear{{Rocha}, {Peter}, {Bullock}, {Kaplinghat},
  {Garrison-Kimmel}, {O{\~n}orbe}  \& {Moustakas}}{{Rocha}
  et~al.}{2013}]{Rocha13}
{Rocha} M.,  {Peter} A. H.~G.,  {Bullock} J.~S.,  {Kaplinghat} M.,
  {Garrison-Kimmel} S.,  {O{\~n}orbe} J.,   {Moustakas} L.~A.,  2013, \mn@doi
  [\mnras] {10.1093/mnras/sts514}, \href
  {https://ui.adsabs.harvard.edu/abs/2013MNRAS.430...81R} {430, 81}

\bibitem[\protect\citeauthoryear{{Rose}, {Torrey}, {Vogelsberger}  \&
  {O'Neil}}{{Rose} et~al.}{2022}]{Rose22}
{Rose} J.~C.,  {Torrey} P.,  {Vogelsberger} M.,   {O'Neil} S.,  2022, \mn@doi
  [\mnras] {10.1093/mnras/stac3634}, \href
  {https://ui.adsabs.harvard.edu/abs/2022MNRAS.tmp.3409R} {}

\bibitem[\protect\citeauthoryear{{Sagunski}, {Gad-Nasr}, {Colquhoun},
  {Robertson}  \& {Tulin}}{{Sagunski} et~al.}{2021}]{Sagunski21}
{Sagunski} L.,  {Gad-Nasr} S.,  {Colquhoun} B.,  {Robertson} A.,   {Tulin} S.,
  2021, \mn@doi [\jcap] {10.1088/1475-7516/2021/01/024}, \href
  {https://ui.adsabs.harvard.edu/abs/2021JCAP...01..024S} {2021, 024}

\bibitem[\protect\citeauthoryear{{Sales}, {Navarro}, {Schaye}, {Dalla Vecchia},
  {Springel}  \& {Booth}}{{Sales} et~al.}{2010}]{Sales10}
{Sales} L.~V.,  {Navarro} J.~F.,  {Schaye} J.,  {Dalla Vecchia} C.,  {Springel}
  V.,   {Booth} C.~M.,  2010, \mn@doi [\mnras]
  {10.1111/j.1365-2966.2010.17391.x}, \href
  {https://ui.adsabs.harvard.edu/abs/2010MNRAS.409.1541S} {409, 1541}

\bibitem[\protect\citeauthoryear{{Sales}, {Wetzel}  \& {Fattahi}}{{Sales}
  et~al.}{2022}]{Sales22}
{Sales} L.~V.,  {Wetzel} A.,   {Fattahi} A.,  2022, \mn@doi [Nature Astronomy]
  {10.1038/s41550-022-01689-w}, \href
  {https://ui.adsabs.harvard.edu/abs/2022NatAs...6..897S} {6, 897}

\bibitem[\protect\citeauthoryear{{Sameie} et~al.,}{{Sameie}
  et~al.}{2021}]{Sameie21}
{Sameie} O.,  et~al., 2021, \mn@doi [\mnras] {10.1093/mnras/stab2173}, \href
  {https://ui.adsabs.harvard.edu/abs/2021MNRAS.507..720S} {507, 720}

\bibitem[\protect\citeauthoryear{{Santos-Santos} et~al.,}{{Santos-Santos}
  et~al.}{2020}]{Santos20}
{Santos-Santos} I. M.~E.,  et~al., 2020, \mn@doi [\mnras]
  {10.1093/mnras/staa1072}, \href
  {https://ui.adsabs.harvard.edu/abs/2020MNRAS.495...58S} {495, 58}

\bibitem[\protect\citeauthoryear{{Schaller} et~al.,}{{Schaller}
  et~al.}{2015}]{Schaller15}
{Schaller} M.,  et~al., 2015, \mn@doi [\mnras] {10.1093/mnras/stv1067}, \href
  {https://ui.adsabs.harvard.edu/abs/2015MNRAS.451.1247S} {451, 1247}

\bibitem[\protect\citeauthoryear{{Schaller} et~al.,}{{Schaller}
  et~al.}{2023}]{Schaller23}
{Schaller} M.,  et~al., 2023, \mn@doi [arXiv e-prints]
  {10.48550/arXiv.2305.13380}, \href
  {https://ui.adsabs.harvard.edu/abs/2023arXiv230513380S} {p. arXiv:2305.13380}

\bibitem[\protect\citeauthoryear{{Schaye} \& {Dalla Vecchia}}{{Schaye} \&
  {Dalla Vecchia}}{2008}]{Schaye08}
{Schaye} J.,  {Dalla Vecchia} C.,  2008, \mn@doi [\mnras]
  {10.1111/j.1365-2966.2007.12639.x}, \href
  {https://ui.adsabs.harvard.edu/abs/2008MNRAS.383.1210S} {383, 1210}

\bibitem[\protect\citeauthoryear{{Schaye} et~al.,}{{Schaye}
  et~al.}{2015}]{Schaye15}
{Schaye} J.,  et~al., 2015, \mn@doi [\mnras] {10.1093/mnras/stu2058}, \href
  {https://ui.adsabs.harvard.edu/abs/2015MNRAS.446..521S} {446, 521}

\bibitem[\protect\citeauthoryear{{Schombert} \& {McGaugh}}{{Schombert} \&
  {McGaugh}}{2014}]{SchombertMcGaugh2014}
{Schombert} J.,  {McGaugh} S.,  2014, \mn@doi [\pasa] {10.1017/pasa.2014.32},
  \href {https://ui.adsabs.harvard.edu/abs/2014PASA...31...36S} {31, e036}

\bibitem[\protect\citeauthoryear{{Shah} \& {Adhikari}}{{Shah} \&
  {Adhikari}}{2023}]{Shah23}
{Shah} N.,  {Adhikari} S.,  2023, \mn@doi [arXiv e-prints]
  {10.48550/arXiv.2308.16342}, \href
  {https://ui.adsabs.harvard.edu/abs/2023arXiv230816342S} {p. arXiv:2308.16342}

\bibitem[\protect\citeauthoryear{{Shen}, {Hopkins}, {Necib}, {Jiang},
  {Boylan-Kolchin}  \& {Wetzel}}{{Shen} et~al.}{2021}]{Shen21}
{Shen} X.,  {Hopkins} P.~F.,  {Necib} L.,  {Jiang} F.,  {Boylan-Kolchin} M.,
  {Wetzel} A.,  2021, \mn@doi [\mnras] {10.1093/mnras/stab2042}, \href
  {https://ui.adsabs.harvard.edu/abs/2021MNRAS.506.4421S} {506, 4421}

\bibitem[\protect\citeauthoryear{{Silverman}, {Bullock}, {Kaplinghat}, {Robles}
   \& {Valli}}{{Silverman} et~al.}{2023}]{Silverman23}
{Silverman} M.,  {Bullock} J.~S.,  {Kaplinghat} M.,  {Robles} V.~H.,   {Valli}
  M.,  2023, \mn@doi [\mnras] {10.1093/mnras/stac3232}, \href
  {https://ui.adsabs.harvard.edu/abs/2023MNRAS.518.2418S} {518, 2418}

\bibitem[\protect\citeauthoryear{{Slone}, {Jiang}, {Lisanti}  \&
  {Kaplinghat}}{{Slone} et~al.}{2023}]{Slone23}
{Slone} O.,  {Jiang} F.,  {Lisanti} M.,   {Kaplinghat} M.,  2023, \mn@doi
  [\prd] {10.1103/PhysRevD.107.043014}, \href
  {https://ui.adsabs.harvard.edu/abs/2023PhRvD.107d3014S} {107, 043014}

\bibitem[\protect\citeauthoryear{{Spergel} \& {Steinhardt}}{{Spergel} \&
  {Steinhardt}}{2000}]{Spergel00}
{Spergel} D.~N.,  {Steinhardt} P.~J.,  2000, \mn@doi [\prl]
  {10.1103/PhysRevLett.84.3760}, \href
  {https://ui.adsabs.harvard.edu/abs/2000PhRvL..84.3760S} {84, 3760}

\bibitem[\protect\citeauthoryear{{Steinmetz} \& {Navarro}}{{Steinmetz} \&
  {Navarro}}{1999}]{Steinmetz99}
{Steinmetz} M.,  {Navarro} J.~F.,  1999, \mn@doi [\apj] {10.1086/306904}, \href
  {https://ui.adsabs.harvard.edu/abs/1999ApJ...513..555S} {513, 555}

\bibitem[\protect\citeauthoryear{{Trayford} et~al.,}{{Trayford}
  et~al.}{2015}]{Trayford15}
{Trayford} J.~W.,  et~al., 2015, \mn@doi [\mnras] {10.1093/mnras/stv1461},
  \href {https://ui.adsabs.harvard.edu/abs/2015MNRAS.452.2879T} {452, 2879}

\bibitem[\protect\citeauthoryear{{Tulin} \& {Yu}}{{Tulin} \&
  {Yu}}{2018}]{Tulin18}
{Tulin} S.,  {Yu} H.-B.,  2018, \mn@doi [\physrep]
  {10.1016/j.physrep.2017.11.004}, \href
  {https://ui.adsabs.harvard.edu/abs/2018PhR...730....1T} {730, 1}

\bibitem[\protect\citeauthoryear{{Tully} \& {Fisher}}{{Tully} \&
  {Fisher}}{1977}]{Tully1977}
{Tully} R.~B.,  {Fisher} J.~R.,  1977, \aap, \href
  {https://ui.adsabs.harvard.edu/abs/1977A&A....54..661T} {54, 661}

\bibitem[\protect\citeauthoryear{{Turner}, {Lovell}, {Zavala}  \&
  {Vogelsberger}}{{Turner} et~al.}{2021}]{Turner21}
{Turner} H.~C.,  {Lovell} M.~R.,  {Zavala} J.,   {Vogelsberger} M.,  2021,
  \mn@doi [\mnras] {10.1093/mnras/stab1725}, \href
  {https://ui.adsabs.harvard.edu/abs/2021MNRAS.505.5327T} {505, 5327}

\bibitem[\protect\citeauthoryear{{Vogelsberger}, {Zavala}  \&
  {Loeb}}{{Vogelsberger} et~al.}{2012}]{Vogelsberger12}
{Vogelsberger} M.,  {Zavala} J.,   {Loeb} A.,  2012, \mn@doi [\mnras]
  {10.1111/j.1365-2966.2012.21182.x}, \href
  {https://ui.adsabs.harvard.edu/abs/2012MNRAS.423.3740V} {423, 3740}

\bibitem[\protect\citeauthoryear{{Vogelsberger}, {Zavala}, {Simpson}  \&
  {Jenkins}}{{Vogelsberger} et~al.}{2014}]{Vogelsberger14}
{Vogelsberger} M.,  {Zavala} J.,  {Simpson} C.,   {Jenkins} A.,  2014, \mn@doi
  [\mnras] {10.1093/mnras/stu1713}, \href
  {https://ui.adsabs.harvard.edu/abs/2014MNRAS.444.3684V} {444, 3684}

\bibitem[\protect\citeauthoryear{{Vogelsberger}, {Zavala}, {Cyr-Racine},
  {Pfrommer}, {Bringmann}  \& {Sigurdson}}{{Vogelsberger}
  et~al.}{2016}]{Vogelsberger16}
{Vogelsberger} M.,  {Zavala} J.,  {Cyr-Racine} F.-Y.,  {Pfrommer} C.,
  {Bringmann} T.,   {Sigurdson} K.,  2016, \mn@doi [\mnras]
  {10.1093/mnras/stw1076}, \href
  {https://ui.adsabs.harvard.edu/abs/2016MNRAS.460.1399V} {460, 1399}

\bibitem[\protect\citeauthoryear{{Vogelsberger}, {Zavala}, {Schutz}  \&
  {Slatyer}}{{Vogelsberger} et~al.}{2019}]{Vogelsberger19}
{Vogelsberger} M.,  {Zavala} J.,  {Schutz} K.,   {Slatyer} T.~R.,  2019,
  \mn@doi [\mnras] {10.1093/mnras/stz340}, \href
  {https://ui.adsabs.harvard.edu/abs/2019MNRAS.484.5437V} {484, 5437}

\bibitem[\protect\citeauthoryear{Walt, Colbert  \& Varoquaux}{Walt
  et~al.}{2011}]{vanderWalt11}
Walt v. d.~S.,  Colbert S.~C.,   Varoquaux G.,  2011, \mn@doi [Computing in
  Science & Engineering] {10.1109/mcse.2011.37}, 13, 22–30

\bibitem[\protect\citeauthoryear{{Wiersma}, {Schaye}, {Theuns}, {Dalla Vecchia}
   \& {Tornatore}}{{Wiersma} et~al.}{2009}]{Wiersma09}
{Wiersma} R. P.~C.,  {Schaye} J.,  {Theuns} T.,  {Dalla Vecchia} C.,
  {Tornatore} L.,  2009, \mn@doi [\mnras] {10.1111/j.1365-2966.2009.15331.x},
  \href {https://ui.adsabs.harvard.edu/abs/2009MNRAS.399..574W} {399, 574}

\bibitem[\protect\citeauthoryear{{Wittman}, {Golovich}  \& {Dawson}}{{Wittman}
  et~al.}{2018}]{Wittman18}
{Wittman} D.,  {Golovich} N.,   {Dawson} W.~A.,  2018, \mn@doi [\apj]
  {10.3847/1538-4357/aaee77}, \href
  {https://ui.adsabs.harvard.edu/abs/2018ApJ...869..104W} {869, 104}

\bibitem[\protect\citeauthoryear{{Yang}, {Nadler}  \& {Yu}}{{Yang}
  et~al.}{2023}]{Yang23}
{Yang} D.,  {Nadler} E.~O.,   {Yu} H.-B.,  2023, \mn@doi [\apj]
  {10.3847/1538-4357/acc73e}, \href
  {https://ui.adsabs.harvard.edu/abs/2023ApJ...949...67Y} {949, 67}

\bibitem[\protect\citeauthoryear{{Ziegler} et~al.,}{{Ziegler}
  et~al.}{2002}]{Ziegler02}
{Ziegler} B.~L.,  et~al., 2002, \mn@doi [\apjl] {10.1086/338962}, \href
  {https://ui.adsabs.harvard.edu/abs/2002ApJ...564L..69Z} {564, L69}

\makeatother
\end{thebibliography}
\bibliographystyle{mnras}

\appendix

\section{Galaxy Stellar Mass Function}\label{AppendixA}

\begin{figure*} 
	\includegraphics[angle=0,width=0.48\textwidth]{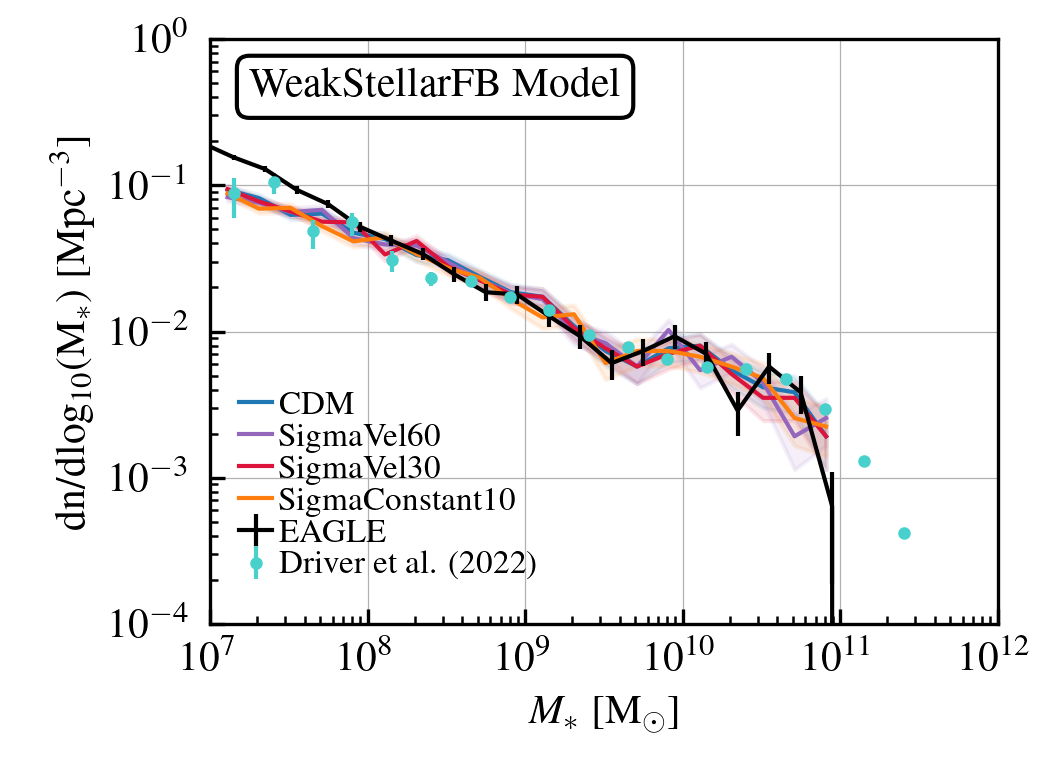}
	\includegraphics[angle=0,width=0.48\textwidth]{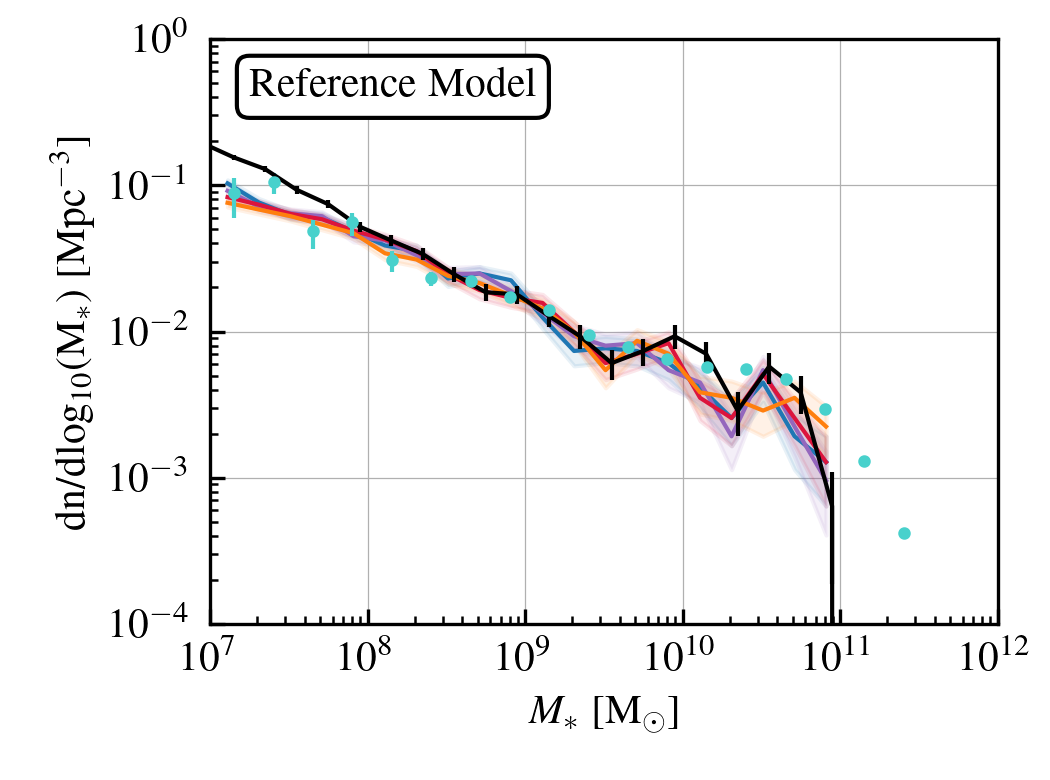}
	\caption{The galaxy stellar mass function at $z=0$ for the `WeakStellarFB' (left panel) and `Reference' (right panel) galaxy formation models under the CDM (blue lines), SigmaVel60 (purple line), SigmaVel30 (red line) and SigmaConstant10 (orange line) schemes. The black line corresponds to the galaxy stellar mass function of the original EAGLE 25 REF model and the green line corresponds the measurements from the DR4 GAMA survey at $z<0.1$ (\citealt{Driver22}). Both models, Reference and WeakStellarFB, produce a galaxy number density in the stellar mass range $10^{8}-10^{11}~\rm{M}_{\odot}$ that is in agreement with EAGLE and the observational data within 0.2 dex. Note that the models considered here were calibrated to reproduce the $z=0$ galaxy stellar mass function.}
	\label{SMF_fig}
\end{figure*}

Fig.~\ref{SMF_fig} shows the $z=0$ galaxy stellar mass function for the WeakStellarFB (left panel) and Reference (right panel) galaxy formation models under the CDM (blue lines), SigmaVel60 (purple line), SigmaVel30 (red line) and SigmaConstant10 (orange line) schemes. The simulation results are compared to the original EAGLE REF model (\citealt{Schaye15}), and to the DR4 Galaxy And Mass Assembly (GAMA) survey (\citealt{Driver22}). The EAGLE data shown throughout this section is taken from the EAGLE reference model run in a (25 Mpc)$^3$ box with the same resolution as the TangoSIDM simulations, which were also run in a (25 Mpc)$^3$ volume.

Both Reference and WeakStellarFB produce a galaxy number density in the stellar mass range $10^{8}-10^{11}~\rm{M}_{\odot}$ that is in close agreement with EAGLE and within 0.2 dex of the observational data. While Fig.~\ref{SMF_fig} seems to indicate that SIDM does not strongly affect the galaxy stellar mass function, it does decrease the number of satellites (as shown in \citealt{Vogelsberger12,Nadler20,Correa22}). SIDM interactions enhance the disruption of subhaloes by tidal stripping from the host. We find that from the 685 satellite galaxies in the stellar mass range $10^{7-10}\rm{M}_{\odot}$ from the Reference/CDM model, 639 (93\%) survive in the Reference/SigmaVel60 model and 544 (79\%) survive in the Reference/SigmaConstant10 model.

\section{Density evolution}\label{AppendixB}

\begin{figure*} 
\includegraphics[angle=0,width=0.7\textwidth]{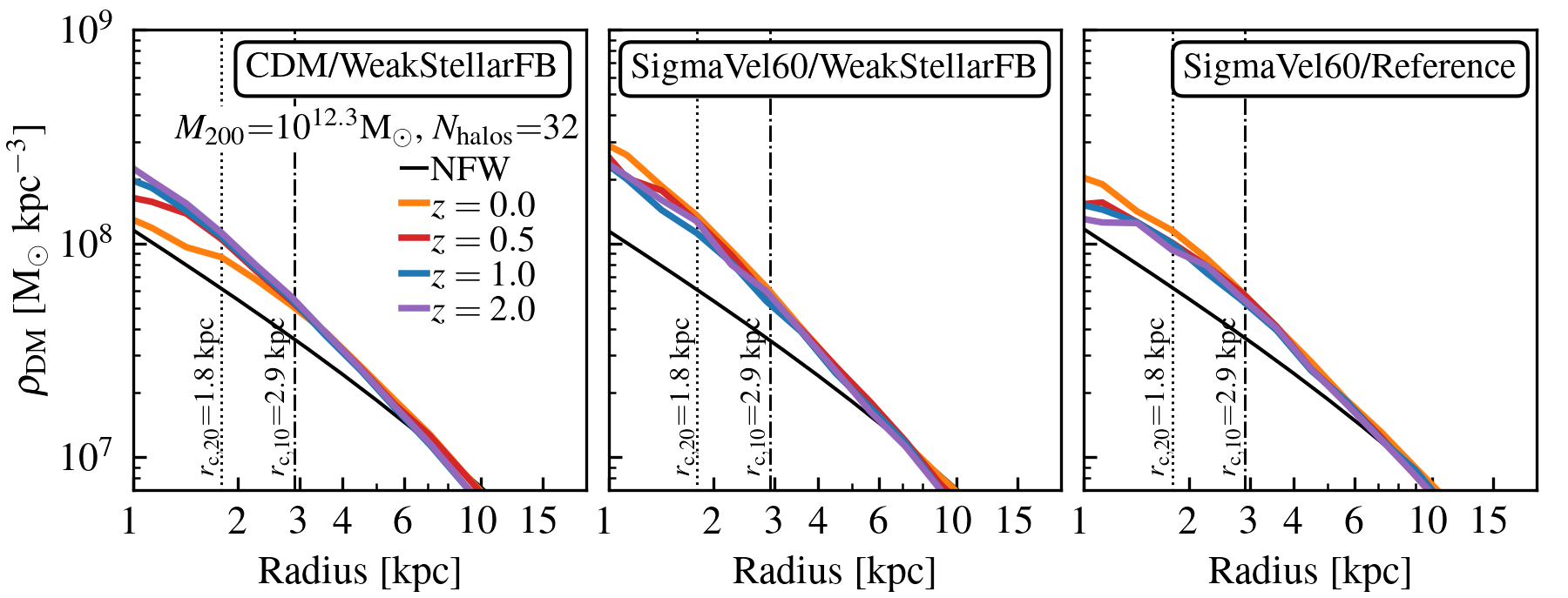}
\caption{Stacked median dark matter density profile, $\rho_{\rm{DM}}$, for the 32 most massive haloes from the WeakStellarFB model under CDM (left panel) and SigmaVel60 (middle panel), and from the Reference model under SigmaVel60 (right panel). The coloured lines show the median values at different redshifts, and the black solid line shows the NFW profile of the haloes at redshift zero. The black dashed-dotted lines indicate the convergence radius (see Section~\ref{Density_sec} for the definition). The median density profiles of haloes over time in WeakStellarFB/CDM are cuspy by redshift 2 and slightly decrease by redshift zero. For the WeakStellarFB/SigmaVel60 model, the median density profiles of the haloes is cuspier by redshift 2, and they do not largely evolve towards redshift zero.}
\label{Density_evolution_2}
\end{figure*}

This appendix expands the discussion presented in Section~\ref{Density_profiles}, where we showed that under SIDM halo dark matter density profiles evolve differently than under CDM (Fig.~\ref{Density_evolution_1}). We have found that as galaxies within SIDM haloes grow in mass, baryons assume a dominant role in the galaxies' central gravitational potential. Consequently, dark matter particles thermalise through frequent interactions, accumulating in the center of the baryon-dominated potential. Fig.~\ref{Density_evolution_2} shows the density evolution of the 32 most massive haloes from the WeakStellarFB model under CDM (left panel) and SigmaVel60 (middle panel), and from the Reference model under SigmaVel60 (right panel). The coloured lines represent the median density evolution between redshifts 0 and 2. In the WeakStellarFB models, the early dominance of baryons in the central potential results in the rapid formation highly cuspy density profiles, which for SIDM remains with minimal evolution in the redshift range zero to two. In contrast, under CDM haloes there is a slight decrease in cuspiness by redshift zero. The right panel of Fig.~\ref{Density_evolution_2} demonstrates that, in the SigmaVel60/Reference model, the median central density of haloes slightly increases over time, as was the case for the SigmaVel30/Reference model.

\section{Assembly history}\label{velocity_sigma_pairs}

Section~\ref{SIDM_param_space} reported an important discrepancy found in massive disc galaxies within the SIDM framework when compared to observations in the Tully-Fisher plane. This discrepancy was translated into an exclusion zone within the SIDM parameter space. Our approach involved identifying velocity-cross section pairs that lead to the formation of galaxies with exceedingly large $V_{\rm{circ}}(R_{\rm{eff}})$. In this section, we provide further details on the methodology employed to determine the lower limits for velocity and cross section, above which the SigmaVel30 and SigmaVel60 models are ruled out.

\begin{figure*} 
\includegraphics[angle=0,width=\textwidth]{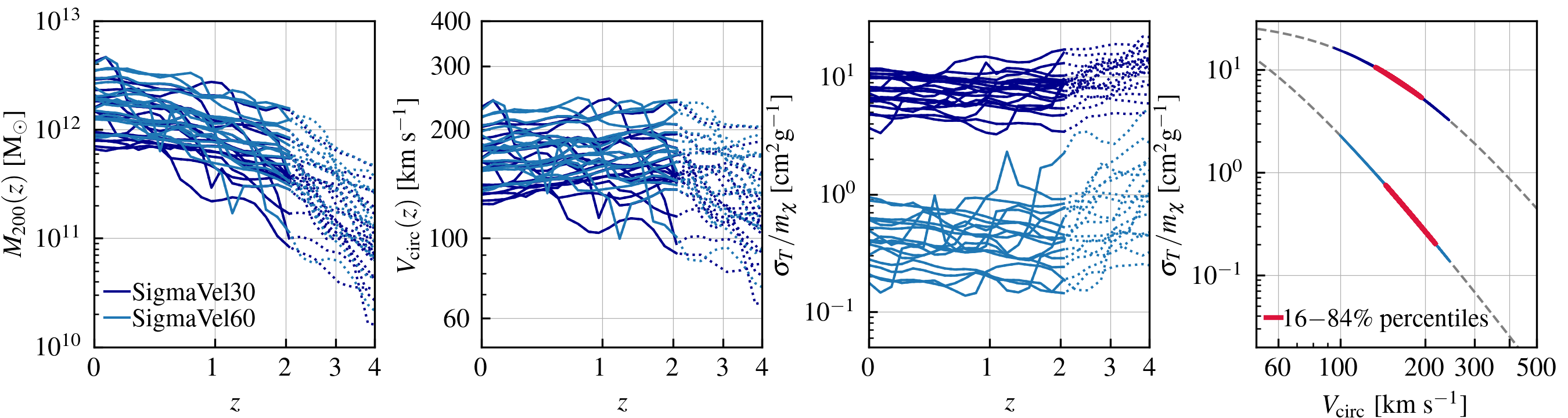}
\caption{{\it{Left panel}}: Mass assembly, $M_{200}(z)$, illustrating the evolution of haloes hosting galaxies that exhibit significant deviations from observations in the $z=0$ Tully-Fisher plane. Dark blue lines represent the mass evolution of individual haloes from the SigmaVel30 model, while light blue lines correspond to the SigmaVel60 model. {\it{Second panel from the left}}: Circular velocity, $V_{\rm{circ}}(z)$, as a function of redshift for the same haloes as in the left panel. {\it{Second panel from the right}}: Cross section, computed from the circular velocity, as a function of redshift. {\it{Right panel}}: Velocity-cross section plane. The grey dashed lines mark the cross section dependence for the SigmaVel30 and SigmaVel60 models. The red limits mark the 16th-84th percentiles of the haloes' evolution in the cross section-velocity plane.}
\label{HaloEvolution}
\end{figure*}

To identify these velocity-cross section pairs, we select all disc galaxies from the Reference + SigmaVel30 and Reference + SigmaVel60 models with stellar masses larger than $10^{10}~\rm{M}_{\odot}$ and $1.3{\times}10^{10}~\rm{M}_{\odot}$, respectively. We follow the assembly histories of the haloes hosting these galaxies across the simulation snapshots until redshift 2 (the redshift below which the haloes' density profiles are well resolved and commence substantial evolution). The left panel of Fig.~\ref{HaloEvolution} shows the mass accretion history $M_{200}(z)$, of the haloes from the SigmaVel30 (dark blue lines) and SigmaVel60 (light blue lines) models under the Reference galaxy model. Converting $M_{200}(z)$ into circular velocity, $V_{\rm{circ}}(z)$, we show these values in the second form the left panel of Fig.~\ref{HaloEvolution}.

We assume that $V_{\rm{circ}}(z)$ corresponds to the average velocity of the dark matter particles within these haloes. Therefore, to estimate the corresponding average dark matter particle cross sections of these haloes, we use eq.~(\ref{sigmam}), assume $v=V_{\rm{circ}}(z)$ and integrate over the scattering angle (as done in eq.~\ref{sigmat}). The evolution of the dark matter haloes' average cross sections, $\sigma_{T}/m_{\chi}$, as a function of redshift is shown in the second panel from the right. The right most panel displays the cross section as a function of the haloes circular velocities for the SigmaVel30 and SigmaVel60 models (grey lines). As expected, all the velocity-cross section pairs that were obtained from the haloes' evolution align with the velocity-$\sigma_{T}/m_{\chi}$ relation from the models (eq.~\ref{sigmat}). 

In the last step, at each redshift we determine the 16-84th percentiles in the distribution of the haloes circular velocities. We highlight these percentage ranges in red in the right panel and mark them as the limits above which the SigmaVel30 and SigmaVel60 models produce overly enhanced central dark matter densities in massive disc galaxies. Therefore, these limits represent the lower bounds above which the SigmaVel30 and SigmaVel60 models are ruled out with $98\%$ and $95\%$ confidence, respectively. 

\begin{table*}
\begin{center}
\caption{Observational data used in this work. Column 2 provides the sample the galaxy belongs to: `S' (SPARC, \citealt{Lelli16}), `R' (\citealt{Reyes11}), `P' (\citealt{Pizagno07}). Note that for the Reyes et al. and Pizagno et al. datasets, the galaxy names correspond to their SDSS names. The complete table can be found online in http://www.tangosidm.com.}
\label{Table_observational_data}
\begin{tabular}{lcrrr}
\hline
Name & Sample & $M_{*}$ [M$_{\odot}$] & $R_{\rm{eff}}$ [kpc] & $V_{\rm{circ}}(R_{\rm{eff}})$ [km s$^{-1}$] \\
\hline
ESO079-G014 &   S  &   2.59e+10 &      7.23 &    140.99  \\
ESO116-G012 &   S  &   2.15e+09 &      2.75 &    80.63  \\
ESO563-G021 &   S  &   1.56e+11 &     10.59 &    294.74  \\
F568-3 &   S  &   4.17e+09 &      7.47 &    91.87  \\
F568-V1 &   S  &   1.91e+09 &      4.40 &    101.01 \\
J001006.61-002609.7 & R & 9.64e+09 &     2.42 &    94.86 \\
J001708.75-005728.9 & R & 4.57e+09 &     3.13 &    107.83 \\
J002844.82+160058.8 & R & 2.91e+10 &     6.08 &    106.65 \\
J003112.09-002426.4 & R & 1.53e+10 &     2.00 &    138.94 \\
J004916.23+154821.0 & R & 7.49e+09 &     5.65 &    107.57 \\
J004935.71+010655.2 & R & 3.72e+10 &     5.40 &    117.47 \\
J011750.26+133026.3 & R & 3.80e+09 &     3.83 &    65.96 \\
J012317.00-005421.6 & R & 1.71e+10 &     2.33 &    137.88 \\
J012340.12+004056.4 & R & 2.14e+10 &     3.01 &    156.31 \\
J012438.08-000346.4 & P & 2.07e+10 &     6.69 &    161.71 \\
J013142.14-005559.9 & P & 6.72e+10 &    13.90 &    225.19 \\
J013600.15+003948.6 & P & 3.10e+10 &     6.15 &    179.73 \\
J013752.69+010234.8 & P & 5.29e+10 &     8.18 &    277.10 \\
J014121.94+002215.7 & P & 1.88e+10 &     3.27 &    195.82 \\
J015746.24-011229.9 & P & 8.53e+10 &     7.19 &    310.80 \\
J015840.93+003145.2 & P & 4.78e+10 &     9.46 &    189.52 \\
... & . & ... & ... & ...\\
\hline
\end{tabular}
\end{center}
\end{table*}

\bsp	
\label{lastpage}
\end{document}